\documentclass[prd,preprintnumbers,superscriptaddress,nofootinbib,twocolumn,balancelastpage,aps,floatfix,notitlepage]{revtex4}

\usepackage[a-2b,mathxmp]{pdfx}

\usepackage{color,graphicx,epsfig,psfrag,amsmath,empheq,amssymb,verbatim,array,layout,textcomp,latexsym}
\usepackage{multirow}
\usepackage{bm}
\usepackage{dcolumn} 
\usepackage{mathrsfs,amsfonts,color}
\usepackage{slashed}
\usepackage{hepunits}
\usepackage{enumitem}
\usepackage{xspace}
\usepackage[utf8]{inputenc}

\newcommand{\cO}{\mathcal{O}}
\newcommand{\cL}{\mathcal{L}}
\newcommand{\cE}{\mathcal{E}}
\newcommand{\cM}{\mathcal{M}}

\newcommand{\cA}{\mathcal{A}}

\newcommand{\cP}{\mathcal{P}}
\newcommand{\cB}{\mathcal{B}}

\newcommand{\eg}{\textit{e.g.\xspace}}

\definecolor{greenp1}{rgb}{0, 0.8, 0}
\definecolor{danielColor}{rgb}{0.9, 0.2, 0.9}

\newcommand{\npulses}{\ensuremath{N_{\rm BX}}\xspace}

\newcommand{\omegaL}{\omega_{\rm L}\xspace}

\newcommand{\Gagam}{\Gamma_\gamma\xspace}
\newcommand{\Gaee}{\Gamma_{ee}\xspace}
\newcommand{\taugam}{\tau_\gamma\xspace}
\newcommand{\tauee}{\tau_{ee}\xspace}

\newcommand{\tL}{t_{\rm L}\xspace}
\newcommand{\nbx}[2]{\ensuremath{\mu_{#1}(#2)}}
\newcommand{\nbxonem}[1]{\ensuremath{\mu_{#1}}}
\newcommand{\fngam}{\ensuremath{f_{n\to\gamma}}\xspace}

\newcommand{\Rsel}{\ensuremath{R_{\rm sel}}\xspace}

\newcommand{\TeV}{{\rm TeV}\xspace}
\newcommand{\GeV}{{\rm GeV}\xspace}
\newcommand{\MeV}{{\rm MeV}\xspace}
\newcommand{\keV}{{\rm keV}\xspace}

\newcommand{\fs}{{\rm fs}\xspace}

\newcommand{\eVol}{\ensuremath{e^-_{\rm V}}\xspace}

\newcommand{\PrimEx}{{\sc PrimEx}\xspace}
\newcommand{\GlueX}{{\sc GlueX}\xspace}
\newcommand{\LUXE}{{\sc LUXE}\xspace}

\newcommand{\Geant}{\textsc{Geant\,4}\xspace}
\newcommand{\MG}{\textsc{MadGraph\,5}\xspace}

\newcommand{\LamUV}{\Lambda_{\rm UV}}

\begin{document}

\title{\LUXE-NPOD: new physics searches with an optical dump at \LUXE}


\author{Zhaoyu Bai}
\email{11712901@mail.sustech.edu.cn}
\affiliation{Department of Physics, Southern University of Science and Technology, Shenzhen 518055, China}
\affiliation{Department of Particle Physics and Astrophysics, Weizmann Institute of Science, Rehovot 7610001, Israel}

\author{Thomas Blackburn}
\email{tom.blackburn@physics.gu.se}
\affiliation{Department of Physics, University of Gothenburg, SE-41296 Gothenburg, Sweden}

\author{Oleksandr Borysov}
\email{oleksandr.borysov@desy.de}
\affiliation{Deutsches Elektronen-Synchrotron DESY, 22607 Hamburg, Germany}

\author{Oz Davidi}
\email{oz.davidi@weizmann.ac.il}
\affiliation{Department of Particle Physics and Astrophysics, Weizmann Institute of Science, Rehovot 7610001, Israel}

\author{Anthony Hartin}
\email{anthony.hartin@desy.de}
\affiliation{University College London, Gower Street, London WC1E 6BT, United Kingdom}

\author{Beate Heinemann}
\email{beate.heinemann@desy.de}
\affiliation{Deutsches Elektronen-Synchrotron DESY, 22607 Hamburg, Germany}
\affiliation{Albert-Ludwigs-Universität Freiburg, 79104 Freiburg, Germany}

\author{Teng Ma}
\email{t.ma@campus.technion.ac.il}
\affiliation{Physics Department, Technion—Israel Institute of Technology, Haifa 3200003, Israel}

\author{Gilad Perez}
\email{gilad.perez@weizmann.ac.il}
\affiliation{Department of Particle Physics and Astrophysics, Weizmann Institute of Science, Rehovot 7610001, Israel}

\author{Arka Santra}
\email{arka.santra@weizmann.ac.il}
\affiliation{Department of Particle Physics and Astrophysics, Weizmann Institute of Science, Rehovot 7610001, Israel}

\author{Yotam Soreq}
\email{soreqy@physics.technion.ac.il}
\affiliation{Physics Department, Technion—Israel Institute of Technology, Haifa 3200003, Israel}

\author{Noam Tal Hod}
\email{noam.hod@weizmann.ac.il}
\affiliation{Department of Particle Physics and Astrophysics, Weizmann Institute of Science, Rehovot 7610001, Israel}

\begin{abstract}
We propose a novel way to search for feebly interacting massive particles, exploiting two properties of systems involving collisions between high energy electrons and intense laser pulses.
The first property is that the electron-intense-laser collision results in a large flux of hard photons, as the laser behaves effectively as a thick medium. 
The second property is that the emitted photons free-stream inside the laser and thus for them the laser behaves effectively as a very thin medium.
Combining these two features implies that the electron-intense-laser collision is an apparatus which can efficiently convert UV electrons to a large flux of hard, co-linear photons.
We further propose to direct this unique large and hard flux of photons onto a  physical dump which in turn is capable of producing feebly interacting massive particles, in a region of parameters that has never been probed before. 
We denote this novel apparatus as ``optical dump'' or NPOD~(new physics search with optical dump).
The proposed \LUXE experiment at Eu.XFEL has all the required basic ingredients of the above experimental concept. 
We discuss how this concept can be realized in practice by adding a detector after the last physical dump of the experiment to reconstruct the two-photon decay product of a new spin-0 particle.
We show that even with a relatively short dump, the search can still be background free.
Remarkably, even with a 40\,TW laser, which corresponds to the initial run, and definitely with a 350\,TW laser, of the main run with one year of data taking,  \LUXE-NPOD will be able to probe uncharted territory of both models of pseudo-scalar and scalar
fields, and in particular probe natural of scalar theories for masses above $100\,\MeV$.
\end{abstract}

\preprint{DESY 21-111}

\maketitle

\section{Introduction}

Despite its great success, the standard model~(SM) of particle physics does not provide a complete description of Nature; new particles and/or forces are required to account for the observed neutrino oscillations, dark matter, and the cosmological baryon asymmetry.
This provides us with a strong motivation to search for new physics~(NP), yet the above observations cannot be robustly linked to a specific microscopic physical scale. 

Therefore, there are worldwide efforts to probe physics beyond the SM~(BSM) at different energy scales in different types of experimental frontiers, see \eg~\cite{Strategy:2019vxc} for a recent discussion. 
Despite all of these efforts we are currently lacking a ``smoking gun'' for a direct observation of NP.
This calls for new experimental approaches that may open the window to alternative ways to search for BSM physics. 

\begin{figure}[t]
	\includegraphics[width=0.45\textwidth]{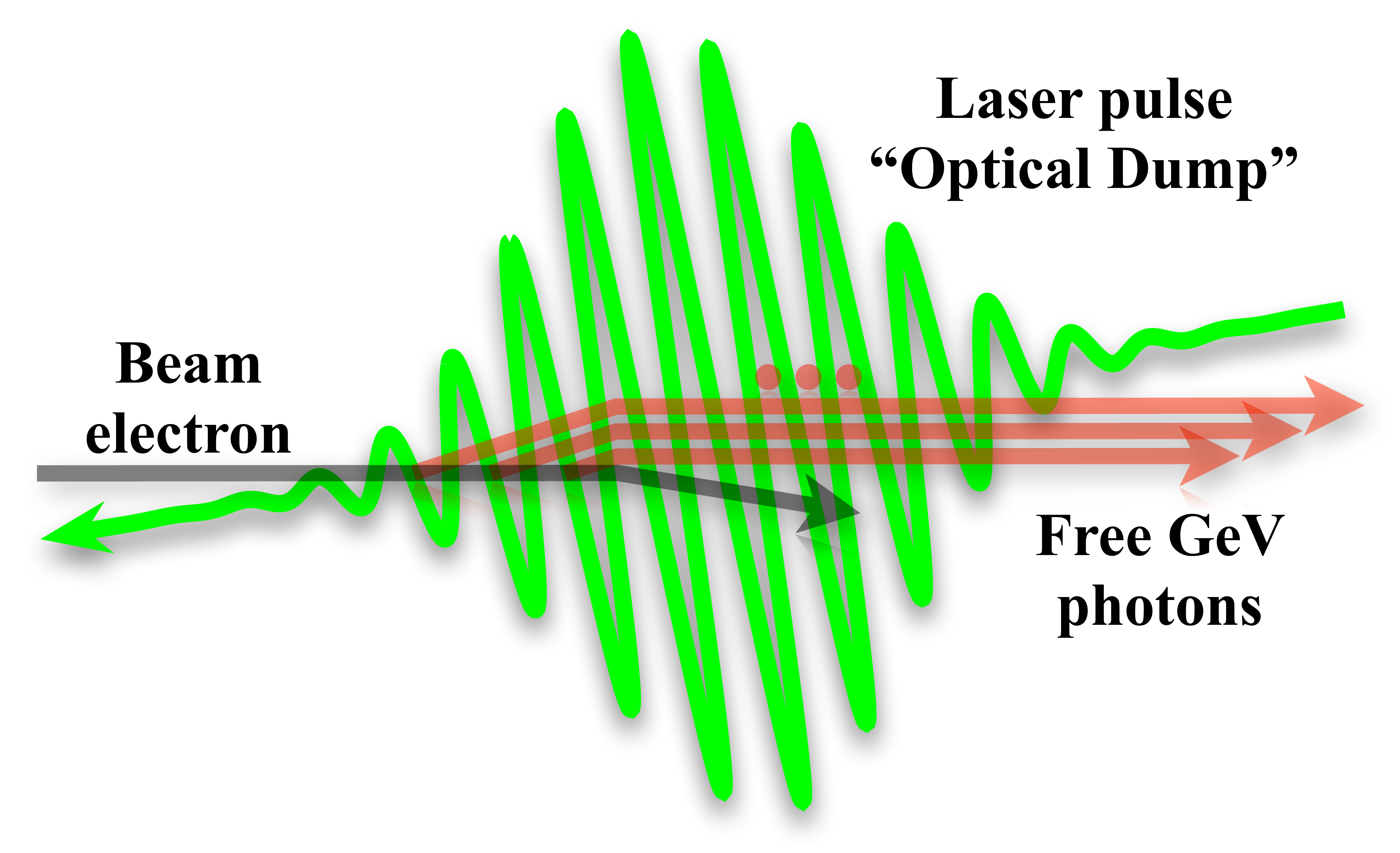}
	\caption{
	Schematic illustration of the optical dump. 
	The high intensity laser pulse behaves effectively as a thick medium for the incoming electron, that in turn may emit a large flux of hard photons which ``free stream'' in this optical medium and can be used to search for new physics.} 
	\label{fig:pulse}
\end{figure}

In this work, we highlight the complementary between the largely unexplored non-perturbative and non-linear quantum electrodynamics~(QED) phenomena, such as Schwinger pair production in a strong electromagnetic~(EM) field~\cite{Sauter:1931zz,Heisenberg:1935qt,Schwinger:1951nm}, and the quest for BSM physics.
We point out that an intense laser behaves as a new type of a particularly-effective dump for the incoming electron beam as follows.   
In the collision between a high-energy electron beam and an intense laser, the laser behaves as a thick medium, leading to the production of a large flux of hard photons~\cite{Ritus1985,DiPiazza:2011tq,Thomas2012,Yan2017,Magnusson:2018sfo}.
Furthermore, as the photons have a negligible interaction with the EM field, they practically free stream in the laser after being produced. 
Thus, the outgoing photon flux can be efficiently used to search for weakly interacting new particles that couples to photons. 
This is illustrated in Fig.~\ref{fig:pulse}.

We show that with an additional forward detector, the proposed \LUXE experiment~\cite{Abramowicz:2019gvx,Abramowicz:2021zja} at the Eu.XFEL~\cite{Altarelli:2006zza} has the potential to probe an unexplored, well motivated, parameter space of new spin-0 (scalar or pseudo-scalar) particles with coupling to photons. 
This proposal is denoted as \LUXE-NPOD: \textit{New Physics at Optical Dump}.
\LUXE is planning to start with a 40\,TW laser and later deploy a 350\,TW laser; 
these two phases are denoted phase-0 and phase-1, respectively.
This setup can probe scalar and pseudo-scalar with masses up to $\cO(350)\,\MeV$ and decay constant of $\cO(10^{5}-10^{6})\,\GeV$, beyond the reach of existing limits.

\section{Electron-laser collisions}

The interaction between high-energy electrons and intense laser pulses is reviewed in ~\cite{Ritus1985,DiPiazza:2011tq,Hartin:2017psj,Gonoskov:2021hwf}.
Here we highlight the relevant points required for the \LUXE-NPOD proposal, focusing on the properties of the radiation generated in the electron-laser collisions at the \LUXE experiment.
The behavior of an electron traveling inside an intense laser pulse can be described by treating the laser as a background field. 
The modified electron-field modes are known as Volkov states~\cite{Volkov}, denoted as $\eVol$ below. 
These states provide an exact solution to the corresponding modified Dirac equation.
Emission processes and pair production are then evaluated using perturbation theory, the ``Furry picutre"~\cite{Furry:1951zz}, similar to what is done in the background-free case. 
In contrast, the photons can be simply described as background-free fields, similar to free photons propagating in space.

The passage of an electron inside the laser pulse is controlled by two processes related to each other by an exchange of the initial and final states.
The first process is the Compton scattering~\cite{Nikishov:1964zza,Brown:1964zzb} (often referred to in the literature as high-intensity or non-linear Compton scattering),
\begin{align}
	\label{eq:HICS}
    \eVol \to \eVol  +\gamma \, ,
\end{align}
where our focus is on cases where, in the lab frame, the electron emits a photon, $\gamma$, at $\cO(\text{few})\,\GeV$.
The typical timescale for this process is $\taugam=1/\Gagam\sim\cO(10)\,\fs$.
The second process to be discussed below, is the Breit-Wheeler pair production~\cite{doi:10.1063/1.1703787} (often referred to in the literature as one photon pair production or non-linear Breit-Wheeler pair production),
\begin{align}
	\label{eq:OPPP}
    \gamma  \to  e^+_{\rm V} + \eVol  \, .
\end{align}
In practice, this process can be viewed as if the original electron first emits a photon, which subsequently interacts with the laser leading to the production of a pair of Volkov states.
The typical pair production timescale relevant to \LUXE is $\tauee=1/\Gaee\sim\cO(10^4-10^6)\,\fs$~\cite{Abramowicz:2021zja}.
The typical laser pulse duration foreseen at \LUXE is $\tL\sim\cO(10-200)\,\fs\,,$ and finally, the relevant time scale of \LUXE's 800\,nm laser itself is $\sim1/\omegaL\sim 0.4\,\fs$~\cite{Abramowicz:2021zja}, where $\omegaL$ is the laser angular frequency.
We find the following hierarchy among these four timescales to be
\begin{align}
    \label{Eq:times}
    1/\omegaL\ll \taugam \lesssim \tL \ll \tauee\,.
\end{align}
Several points are in order as follows:
(i)~the fact that $1/\omegaL$ is the shortest scale in the problem supports the treatment of the laser as a background field to a leading order;
(ii)~the fact that $\tauee$ is much longer than all the other scales in the problem implies that we can treat the photons in the laser as free streaming;
(iii)~the fact that $\taugam$ is shorter than $\tL$ implies that it behaves as a thick target for the electrons. 
For an ideal large pulse (spatially and temporally), the electrons will in principle lose all their energy to the photons. 

The combination of points (ii) and (iii) above and the resulting hard spectrum of photons is, as already mentioned above, the core reason for why we denote our experiment as optical dump and why we believe it provides us with a novel concept to search for feebly interacting massive particles. 

In practice, due to the limited size and duration of the laser's pulse~\cite{Abramowicz:2021zja}, the beam electrons at \LUXE are not  stopped.
These electrons are deflected away by a magnet right after the interaction with the laser.
The region after the electron-laser interaction chamber in Fig.~\ref{fig:scheme} can be considered as being effectively free from any electrons that passed the optical dump region.

To contrast the optical dump with a conventional solid-dump, consider the propagation of high energy electron or photon in the dump. 
In both cases the mean free path is controlled by scattering of the highly charge heavy nucleus, which is of the order of the electron radiation length, $X_0$ see \eg~\cite{Zyla:2020zbs}. 
This is an important difference between the laser medium and a solid-material medium.
In the former, the pulse length can be made long compared to the photon production timescale, while being short enough compared to the pair production one, i.e.  $\taugam \lesssim \tL \ll \tauee$.
Therefore, a few hard photons (with $E_\gamma \gtrsim 1\,\GeV$ on average) per incoming electron exit the laser pulse.
In the thick limit of a solid-material dump, if the material length, $d$, is much larger than $X_0$, all of the hard photons will be absorbed in the material. 
In the thin solid-material limit, where $d\ll X_0$, the hard Bremsstrahlung photons can escape the material, but their production rate is suppressed by $d/X_0\ll 1$.
For example, in phase-1 of \LUXE\ we expect $\gtrsim 3.5$ photons per incoming electron inclusively (and $\gtrsim 1.7$ photons with $E_\gamma>1\,\GeV$ per incoming electron), while in the thin solid-material limit, much less photons are expected (\eg{} for $d/X_0\sim0.1$ only $\sim0.06$ photons with $E_\gamma>1\,\GeV$ are emitted per incoming electron).

\section{New physics scenarios}

In this work, we mostly focus on new spin-0 particles, which are found in many well motivated extensions of the SM.
We focus on two cases of a pseudo-scalar, $a$, which is often denoted generically as an axion-like-particle~(ALP), and of a scalar, $\phi$.
Light ALPs arise in variety of models motivated by the Goldstone theorem, with their masses protected by a shift symmetry, see~\cite{Marsh:2015xka,Graham:2015ouw,Hook:2018dlk,Irastorza:2018dyq,Choi:2020rgn} for recent reviews.
In addition we consider a CP even, $\phi$, where theoretically, constructing a natural model of a light scalar is rather challenging. However, two concrete proposals have been put forward, one where the scalar mass is protected by an approximate scale-invariance symmetry~(see for instance~\cite{Goldberger:2007zk} and Refs. therin), and a second one where it is protected by an approximate shift-symmetry that is broken, together with CP~\cite{Flacke:2016szy, Choi:2016luu}, by two sequestered sectors (inspired by the relaxion paradigm~\cite{Graham:2015cka}). 
Below, we for simplicity consider CP conserving model, described by a single coupling, thus, we consider either ALP or scalar models. 

Our main focus in this work would be on models with effective ALP and scalar coupling to photons,
\begin{align}
	\label{eq:Laphi}
	\cL_{a,\phi} 
=	\frac{a}{4\Lambda_a} F_{\mu\nu}\tilde{F}^{\mu\nu} 
	+\frac{\phi}{4\Lambda_\phi} F_{\mu\nu}{F}^{\mu\nu} \, , 
\end{align}
where $\tilde{F}_{\mu\nu} = \frac{1}{2}\epsilon_{\mu\nu\alpha\beta}F^{\alpha\beta}$.
Since the ALP is a pseudo-Goldstone mode, its mass, $m_a$, can be much smaller than the scale of its interaction with SM particles, i.e. $\Lambda_a \gg m_a$. 
The decay rate of the ALP and scalar into two photons are given by $\Gamma_{a\to2\gamma} = m^3_a/(64\pi\Lambda_a^2)$ and $\Gamma_{\phi\to2\gamma} = m^3_\phi/(64\pi\Lambda_\phi^2)$, respectively.

The $\phi$-photons coupling induces, quadratically divergent, additive contribution to the scalar mass-square, $\delta m_\phi^2\sim \LamUV^4/(16 \pi^2\Lambda_\phi^2)$ which leads to a naturalness bound
\begin{align}
	\label{eq:natu}
    	\Lambda_\phi
    	\gtrsim  
    	4\times 10^5\, \GeV 
    	\left(\frac{\LamUV}{\TeV}\right)^2 
    	\frac{200\,\MeV}{m_\phi} \, ,
\end{align}
with $\LamUV$ is the scale in which NP is required to appear in order to cancel the quadratic divergences.
Below, we show that the \LUXE-NPOD experiment is expected to reach the sensitivity required to probe the edge of the parameter space of natural models in its phase-1. 
Moreover, the same loop diagram as above induces a mixing between the Higgs and the $\phi$ scalar.
This mixing can be estimated by calculating the square mixed mass term $\delta \mu^2_{H\phi}\sim \LamUV^4\alpha /(64 \pi^3\Lambda_\phi v)$, where $v\simeq246\,\GeV$ is the Higgs VEV.
Thus, the $H-\phi$ mixing is
\begin{align}
    \theta_{H\phi}
    \sim 
    10^{-6} 
    \left( \frac{\LamUV}{\TeV} \right)^4
     \frac{4\times 10^5\,\GeV}{\Lambda_\phi}  \, ,
\end{align}
which is in an unconstrained region of Higgs portal (or relaxion) models' parameter space. See~\cite{Strategy:2019vxc,Banerjee:2020kww} for a recent analysis.

Alternatively, one can match the coupling $\Lambda_\phi$ in Eq.~\eqref{eq:Laphi} to models of Higgs-scalar mixing (or the relaxion~\cite{Graham:2015cka,Flacke:2016szy,Choi:2016luu}), which leads to $\Lambda_\phi\sim 4\pi v/(\alpha\sin\theta_{H\phi})$. 
In case of inflation based relaxion models one finds $\sin \theta_{H\phi}\lesssim (m_\phi/v)^{2/3}$~\cite{Banerjee:2020kww}, implying $\Lambda_\phi\gtrsim 5\times 10^7\,\GeV (200\,\MeV/m_\phi)^{2/3}\,$ which is beyond the reach of \LUXE-NPOD.

We briefly explore the ALP/scalar electron coupling, which is capture by 
\begin{align}
	\label{eq:Laphie}
	\cL_{e,a,\phi}
=	ig_{ae}a \bar{e}\gamma^5 e + g_{\phi e} \phi \bar{e} e \, .
\end{align}

Finally, we also consider ``milli-charged'' particles~(mCP)~\cite{Holdom:1985ag,Dienes:1996zr,Abel:2003ue,Batell:2005wa}, denoted here as $\psi$, with a mass $m_\psi\ll m_e$ and a fractional electric charge $q\ll 1$.
The effective mCP-photon interaction can be simply written as
\begin{align}
	\label{eq:Lpsi}
	\cL_{\psi} = e q \bar{\psi}\slashed{A} \psi \, .
\end{align}
%

\section{New physics production}

In this section with discuss the NP production mechanisms.
We mainly focus on processes involving ALP and scalar, which are similar qualitatively and quantitatively, we therefore denote them simply as $X=a,\phi\,.$

The first mechanism, which is also the main focus of this work, is \textit{Secondary NP production}.
In this case, the photons produced in the electron-beam and laser pulse collisions (see Eq.~\eqref{eq:HICS}), are freely propagating to collide with the nuclei, $N$, of the material of some sizable dump to produce NP.
In this case we focus on the Primakoff production of $X$'s
\begin{align}
    \label{eq:Primakoff}
	\gamma + N \to  N + X \, .
\end{align}

The second mechanism is \textit{Primary NP production}, where the NP is directly produced at the electron-laser interaction region via $X$ electron coupling in Eq.~\eqref{eq:Laphie}. 
In analogy to Eq.~\eqref{eq:HICS}, we consider the $X$ version of the process.
In the case where $X$ has couplings to electron only, one gets
\begin{align}
	\label{eq:aComptonProd}
    	\eVol \to \eVol + X \, , 
\end{align}
This case is studied in~\cite{Dillon:2018ouq,Dillon:2018ypt,King:2018qbq,King:2019cpj}.
In the case where $X$ has couplings to photons only, one gets
\begin{align}
	\label{eq:aOffshellProd}
	\eVol \to 
	\eVol + \gamma^* \to \eVol + \gamma + X \, . 
\end{align} 
In analogy to Eq.~\eqref{eq:OPPP}, we can also consider the direct production of mCP pairs
\begin{align}
	\label{eq:mCPprod}
	\gamma\,({\rm or}\,\,\gamma^*)  \to \psi^+ + \psi^-.
\end{align}
For charged scalar pair production see~\cite{VillalbaChavez:2012bb}.

While in the primary production case the new particle mass is limited to $m_{X,\psi} \lesssim \cO(100)\,\keV$ (for detail see the Supplemental Material), above this mass scale the production rate becomes negligibly small), in the secondary production case an ALP with a mass up to $\cO(1)\,\GeV$ can be produced in a coherent Primakoff production.

\begin{figure}[t]
	\includegraphics[width=0.45\textwidth]{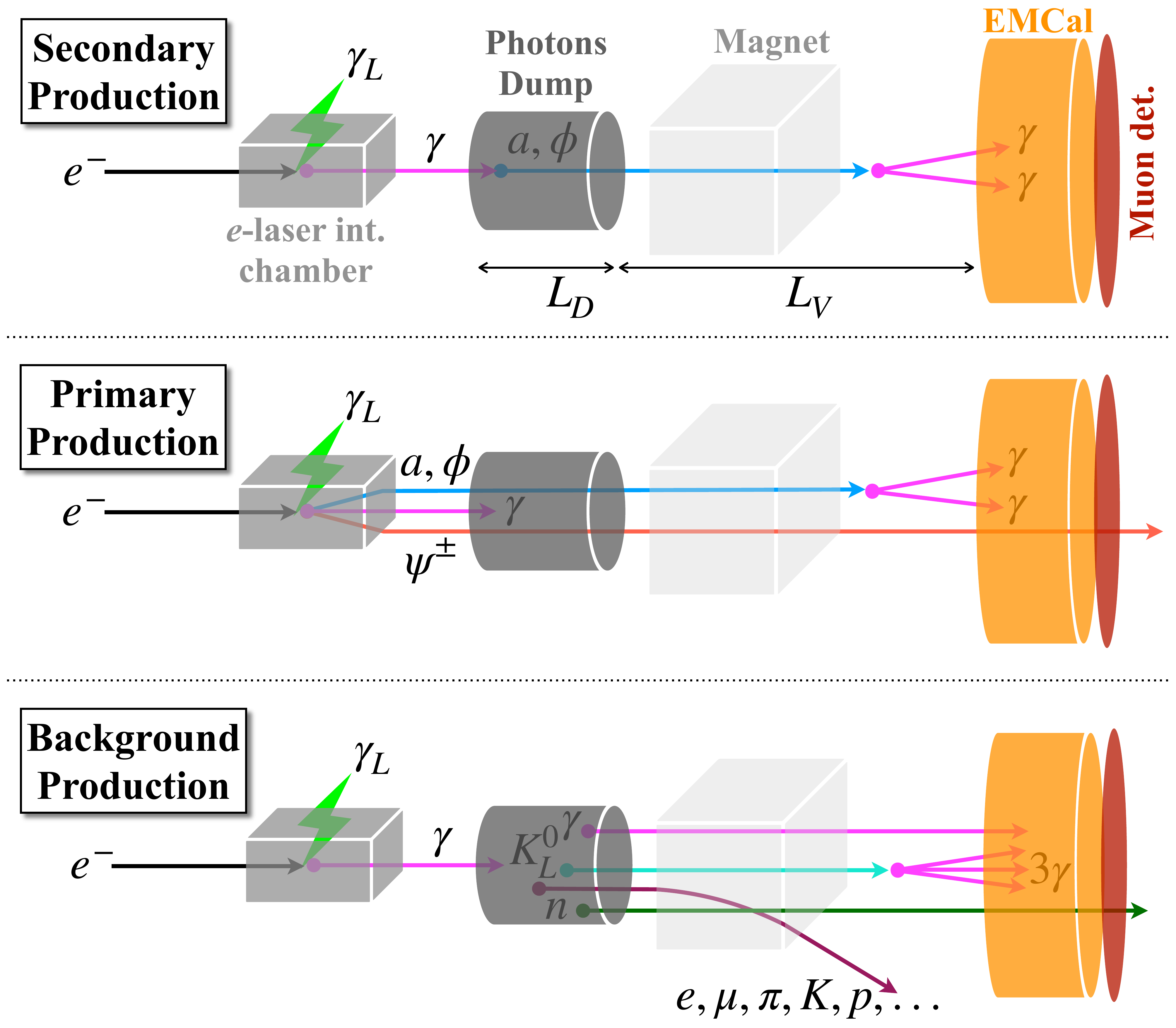}
	\caption{
	An illustration of the \LUXE-NPOD concept and the different search mode. 
	Shown are schematics of the secondary (top) and primary (middle) production mechanisms realisation in the experimental setup. 
	The relevant background topologies are also shown (bottom). 
	The electrons are deflected by a magnet placed right after the interaction chamber.
	}
	\label{fig:scheme}
\end{figure}

An illustration of these two possibilities is provided in Fig.~\ref{fig:scheme}, in the top and middle pannels, in the context of \LUXE along with the associated background topologies, in the bottom pannel.

In the following discussion, we focus only on a detailed feasibility study for the secondary NP production in the context of \LUXE.
The expected signal rates for the primary NP production in Eqs.~\eqref{eq:aComptonProd}--\eqref{eq:mCPprod} are briefly discussed in the Supplementary Material and we leave the detailed study for future work. 

\section{The LUXE-NPOD proposal}
\label{sec:npodproposal}

We have arrived to the heart of this work where we propose to use the high flux of GeV photons, emitted from \LUXE's electron-laser interaction region, to search for feebly interacting spin-0 particles. 
The properties of the photon-flux are described in detail below, but to complete the picture of our main search mode, we discuss the experimental setup assuming a given flux of hard photons. 
After being produced, the photons freely propagate to a physical dump and interact with its nuclei to produce the $X$ particle.

The dump is of length $L_D$ and it is positioned $\sim 13$\,m  away from the electron-laser interaction region.
The $X$ particles are long-lived and hence, they will travel some distance before decaying into $\gamma\gamma$.
Therefore, an empty volume of length $L_V$ is left at the back of the dump to allow the $X$ particles to decay back into two photons.
The two-photon signature is our signal in the detector which is positioned at $L_V$ after the dump.

The expected number of $X$'s produced and detected in the proposed setup shown in Fig.~\ref{fig:scheme} (middle) can be approximated as (see  \eg~\cite{Berlin:2018pwi,Chen:2017awl}),
\begin{align}
	\label{eq:NaSec}
	N_{X}
	\approx
	\cL_{\rm eff}\!\!
	\int \! d E_{\gamma}
	\frac{d N_{\gamma}}{dE_{\gamma}}  \sigma_{X}(E_{\gamma})
	\!\left( e^{-\frac{L_D}{L_{X}}} \!-\! e^{-\frac{ L_V+L_D}{L_{X}}} \right) 
	\cA \, ,
\end{align}
with $\sigma_X(E_{\gamma})$ is the Primakoff cross section that depends on the square of the nuclear charge $Z^2$, $\cL_{\rm eff}$ is an effective luminosity 
term discussed further below, $E_{\gamma}$ is the incoming photon energy, $L_{X} \equiv  c \tau_{X} p_{X}/m_{X} $ is the propagation length of the $X$ particle, with $\tau_{X}$ and $p_{X} \approx \sqrt{E_{\gamma}^2-m^2_{X}}$ being its proper life-time and momentum, respectively. $\sigma_{X} (E_{\gamma})$ is the Primakoff production cross section of $X$ (see \eg~\cite{Tsai:1986tx,Aloni:2019ruo}), and $\cA$ is the angular acceptance and efficiency of the detector.

We estimate $N_X$ independently by using a \MG~v\,2.8.1~\cite{Stelzer:1994ta,Alwall:2011uj,Alwall:2014hca,Degrande:2011ua} Monte Carlo simulation of the process shown in Eq.~\eqref{eq:Primakoff}, including an event-by-event acceptance estimation. 
We use the UFO model~\cite{Degrande:2011ua} from Ref.~\cite{Brivio:2017ije} and follow Refs.~\cite{Tsai:1973py,Bjorken:2009mm,Chen:2017awl} for the form factors.
The decay of $X$ in this \MG{} simulation is instantaneous.
Therefore, to simulate the distance which the $X$ travels before it decays, a random length parameter is drawn from the exponential distribution defined by the particular $X$ propagation length as discussed above.
The decay point is obtained by displacing the $X$ particle from its production point by that random distance according to the 3D direction dictated by its momentum at the production point.
The results of the approximation in Eq.~\eqref{eq:NaSec} and the \MG{} simulation are found to be in a very good agreement.

As a benchmark, we consider a tungsten~($W$) dump with $L_D=1.0$\,m and a radius of at least $\sim 10$\,cm.
With this choice, the effective luminosity can be written as $\cL_{\rm eff}=N_e \npulses \frac{9\rho_W X_0}{7A_W m_0}$~\cite{Tsai:1973py}, where $\rho_W$ is the $W$ density, $A_W$ is its mass number and $X_0$ is its radiation length (all taken from~\cite{Zyla:2020zbs}). 
The remaining parameter is the nucleon mass, $m_0=1.66 \times 10^{-24}\,$g ($\sim 930\,$MeV).
An electron beam with a bunch population of $N_e=1.5\times 10^9$ electrons is assumed with a fixed energy of $E_e=16.5\,\GeV$, and the outgoing photon spectrum from the electron-laser interaction is denoted as $dN_\gamma/dE_{\gamma}$.
These are typical Eu.XFEL operation parameters as discussed in~\cite{Abramowicz:2021zja}.
We further assume one year of data taking.
This duration corresponds to $10^7$ effective live seconds of the experiment and thus, actual $\npulses=10^7$ laser-pulse and electron-bunch collisions (bunch-crossings, BX) in one year\footnote{The collision rate is mostly limited by the laser repetition rate at about 10\,Hz. \LUXE is expecting to operate at 10\,Hz, broken into a collision rate of 1\,Hz plus a rate of 9\,Hz without laser pulses to study the background induced by the electron beam itself in the other parts of the experiment.}.

For the detector, we assume a disk-like structure with a radius of $R=1$\,m positioned at $L_V=2.5$\,m after the end of the dump and concentric with it.
We assume a minimal photon-energy threshold of $0.5\,\GeV$ for detection and hence, a signal event is initially identified as having two photons in the detector surface with $E_\gamma>0.5\,\GeV$. 
Further requirements are discussed below.

The differential photon flux per initial electron, $dN_{\gamma}/dE_{\gamma}$, includes  photons from the electron-laser interaction, as well as secondary photons produced in the EM shower which develops in the dump.
With $L_D\gtrsim 0.2$\,m, all of the primary photons are stopped in the dump.

\begin{figure}[t]
	\includegraphics[width=0.45\textwidth]{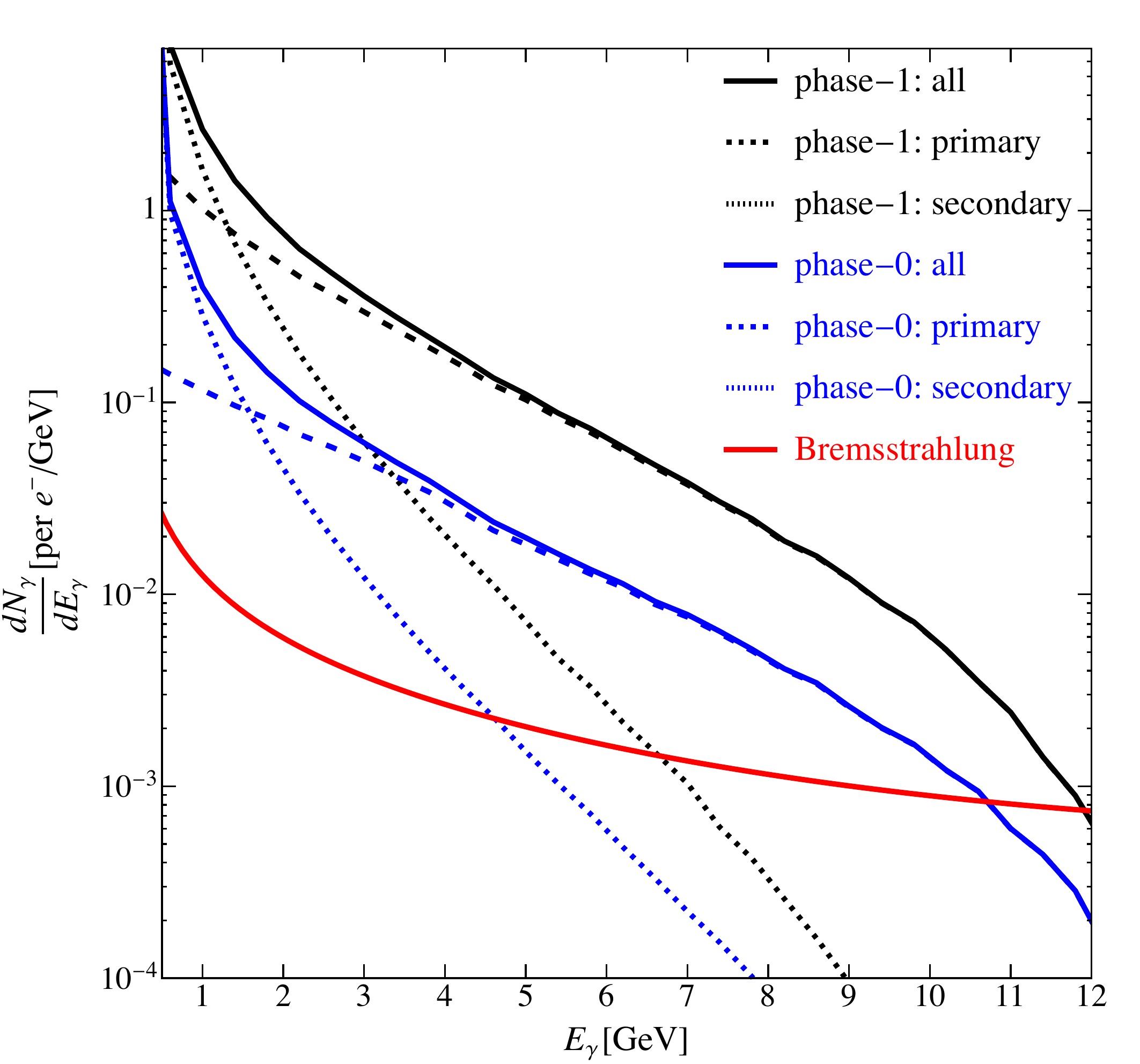}
	\caption{The emitted photon spectrum for phase-0\,(1) in blue\,(black) compared to the perturbative 
	Bremsstrahlung spectrum with $E_e=16.5\,\GeV$ and target length of $0.01X_0$ in red. }
	\label{fig:dNdE}
\end{figure}

The primary photon flux is determined from full strong-field QED Monte Carlo simulation~\cite{ptarmigan,Blackburn:2021rqm}, using probability rates derived in~\cite{Heinzl:2020ynb} and corresponding to the two \LUXE benchmarks (phase-0 and phase-1) which mostly differ by the laser power (40\,TW and 350\,TW respectively).
We assume a laser pulse length of $25\,(120)\,\fs$ for phase-0\,(1) with a transverse spot-size of $6.5\,(10)\,\mu$m respectively.
These pulse configurations correspond to an intensity parameter of $\xi\simeq 3.2\,(3.4)$ gp{Say it better correpond to phases} for the two phases, respectively.
The laser intensity parameter is $\xi\equiv e\cE/(m_e\omegaL)$ where $\cE$ is magnitude of the laser's electric field and $\omegaL\sim1.5$\,eV is the energy of the laser photon (at 800\,nm wavelength).
The sensitivity of the experiment for $X$ searches can be maximized by optimizing the pulse length with respect to the spot-size (within their reasonable ranges) such that the flux of photons with $E_\gamma \gtrsim 1\,\GeV$ will be maximal.
For a dump at a distance of $13$\,m from the interaction point, about 95\% of the emitted photons will fall inside a radius of 5\,cm. 
The  photon passage through the dump material and the evolution of the shower in it are simulated with \Geant v~10.06.p01~\cite{Agostinelli:2002hh,Allison:2006ve,Allison:2016lfl}. 
The resulting (primary and secondary) photons spectra is shown in Fig.~\ref{fig:dNdE}.
Since the emitted Compton photon flux was never measured at the \LUXE\ laser parameters, we propose to use the measured flux in order to normalize $N_a$ in-situ, i.e. taking $dN_{\gamma}/dE_{\gamma}$ from data.

As discussed above, a tungsten dump of $L_D=1.0$\,m is effectively blocking all incident primary Compton photons.
However, the SM particles produced in the dump during the shower generate backgrounds of three types:
(i) charged particles, namely electrons, muons and hadrons;
(ii) fake photons: mostly neutrons misidentified as photons;
(iii) real photons: mostly from EM/hadronic interactions close to the end of the dump, or from meson decays in the volume.

The rates of these different background components are estimated using a detailed \Geant simulation in the same way as discussed above.
While the background levels for phase-0 are softer and smaller than those of phase-1, the same levels are conservatively assumed hereafter also for phase-0.

The rate of charged SM particles (mostly muons and protons), which in principle arrive at the detector with a minimum energy of 0.5\,GeV, is smaller by roughly a factor of 10 compared to the rate of neutrons with the same characteristics (the neutrons are also typically harder).
These particles can be effectively bent away from the detector surface by a magnetic field of $B\approx 1$\,T over an active bending length of $\sim 1$\,m.
Furthermore, muons from the dump or from cosmic rays, which do arrive at the detector can be vetoed with dedicated muon-chambers placed near (mostly behind) the photon detector.
Hence, the charged particle component is not considered as background in the following discussion.

Thus, focusing on background photons and/or neutrons, we denote the average number of particles with energy above 0.5\,GeV arriving at the detector surface per one BX as \nbx{x}{L_D}.
In this notation, $x$ is a neutron ($x=n$) or a photon ($x=\gamma$) and $L_D$ is measured in meters.

Following a \Geant run with $10^{10}$ primary photons (equivalent to $\sim 2$ bunch crossings) which are distributed according to the photon spectrum of \LUXE's phase-1 (see Fig.~\ref{fig:dNdE}), we find that the number of such neutrons per BX is $\nbx{n}{1.0\,\mbox{m}}=10\pm2.3$, where the error is statistical only.
In the same run, we also find zero photons and thus, we can only infer from it that $\nbx{\gamma}{1.0}\ll1$.
A statistically precise estimation of \nbx{\gamma}{1.0\,\mbox{m}} is rather challenging computationally, since the number of BXs simulated has to be a few orders of magnitude larger than what is simulated now. 

We approach the problem of extracting \nbx{\gamma}{1.0\,\mbox{m}} 
in the following two independent analyses that yield consistent results. 
Both approaches are based on the fact that we model the amount of particles that exit the dump as a function of its length, by fitting the results from repeated simulation runs for different $L_D$ values below its nominal value to allow for adequate background photon statistics.
For $L_D<1$\,m, both \nbx{\gamma}{L_D} (and \nbx{n}{L_D}) becomes larger such that our statistical error becomes sufficiently small and we can confidently fit the model's parameters.
In our first analysis we simply model the exiting photon flux as an exponentially falling distribution of $L_D$, while in the second we assume that the photon to neutron number ratio is constant with $L_D$.
We discuss below in more detail the second approach. However, both approaches yield a consistent result as expected, since both background sources are dominated by the hadronic activity close to the end of the dump, with photons mostly originating from meson decay near the dump-edge.
This correlation between the photons and neutrons production in the dump as well as the assumption that the ratio is approximately constant are briefly demonstrated in the Supplemental Material.

In the second approach, the extrapolation to the nominal case of $L_D=1.0$\,m is done by fitting the ratio $R_{\gamma/n}=\nbx{\gamma}{L_D}/\nbx{n}{L_D}$ vs $L_D<1.0$\,m to a zeroth order polynomial and multiplying the result by the number of neutrons per BX obtained for the nominal case of $L_D=1.0$\,m, i.e. $\nbx{\gamma}{1.0\,\mbox{m}}\approx\nbx{n}{1.0\,\mbox{m}} \times R_{\gamma/n}\,$.
The result of the fit for five such runs starting from $L_D=0.30$\,m and going up to $L_D=0.50$\,m in steps of $0.05$\,m is $R_{\gamma/n}=0.0013\pm 0.0002$.
The reduced $\chi^2$ of the fit is 1.88.
The data and fit can be seen in the Supplemental Material.
The extrapolated number of photons per BX for the nominal case of $L_D=1.0$\,m is therefore $\nbx{\gamma}{1.0\,\mbox{m}}=0.013\pm 0.004$.
In the following discussion we will omit the dump length notation, while still assuming $L_D=1.0$\,m, i.e. $\nbxonem{\gamma}=\nbx{\gamma}{1.0\,\mbox{m}}$ and $\nbxonem{n}=\nbx{n}{1.0\,\mbox{m}}$

The number of background events over some period of run-time, where two photons are detected in the same BX can be calculated from the probability to find two real photons or two fake photons (neutrons misidentified as photons) or one real photon and one fake photon per BX in the detector volume:
\begin{itemize}
    \item the probability to find two real photons is $P_{2\gamma} = \cP(\nbxonem{\gamma},2)$, where $\cP$ is a Poisson probability, 
    \item the probability to find two fake photons from neutrons is $P_{2n\to2\gamma} = \sum_{k_n=2}^{\infty} \cP(\nbxonem{n},k_n)\times\cB(2,k_n,\fngam) = \fngam^2 e^{-\nbxonem{n}\fngam} \nbxonem{n}^2/2$, where $\cB$ is a binomial probability, $k_n$ is the number of neutrons and \fngam\ is the probability to misidentify a neutron as a photon with $E_\gamma > 0.5$\,GeV, and
    \item the probability to find one real photon and one fake photon from a neutron is $P_{n+\gamma\to2\gamma} = \cP(\nbxonem{\gamma},1)\times \sum_{k_n=1}^{\infty} \cP(\nbxonem{n},k_n)\times\cB(1,k_n,\fngam) = \left(\nbxonem{\gamma}e^{-\nbxonem{\gamma}}\right)\left(\fngam e^{-\nbxonem{n}\fngam} \nbxonem{n}\right)$.
\end{itemize}
For the values of \nbxonem{\gamma} and \nbxonem{n} obtained above, the resulting probabilities are
$P_{2\gamma} \approx 8.3\times 10^{-5}$, $P_{2n\to2\gamma} \approx 50 \fngam^2 e^{-10\fngam}$ and
$P_{n+\gamma\to2\gamma} \approx 0.13 \fngam e^{-10\fngam}$,
respectively.
The ``fake rate'', \fngam, depends strongly on the specific detector technology choice which has to be made such that \fngam\ is small enough.
The dump itself can be further optimized to maximize the effective luminosity and the signal production rate, while minimizing the hadronic interaction length, using a combination of materials, and an improved dump and detector geometry.

Besides minimizing \fngam, the number of two-photon background events estimated from the probabilities above can be reduced by a set of selection requirements based on the reconstructed properties of the two-photon system. 
These may include, for example, requirements on the invariant mass, the common vertex, the $X$ production vertex and the timing. 
As for \fngam, the projected performance of these requirements depends strongly on the detector technology.
The rejection power of the full selection criteria per BX is hereafter denoted as $\Rsel$.
For simplicity, we assume that $\Rsel$ is similar between the different background components.
The number of background two-photon events in one year (with $\npulses=10^7$) is estimated to be $N_{\rm b}=\npulses\, P_{\rm b}\,\Rsel$, where ${\rm b}=2\gamma,\,2n\to2\gamma$ and $n+\gamma\to2\gamma$ for the three background components respectively.
The numerical estimates for these three channels are therefore:
\begin{align}
	N_{2\gamma}
&	\approx 
	8.3\times 10^2 \, \Rsel  \, , \\
	N_{2n\to2\gamma}
&	\approx
	5.0\times 10^8 \, \fngam^2 e^{-10\fngam} \, \Rsel \,, \\
	N_{n+\gamma\to2\gamma}
&	\approx
	1.3\times 10^6 \, \fngam e^{-10\fngam} \, \Rsel \,.
\end{align}
We see that with $\Rsel \lesssim 10^{-3}$ and $\fngam \lesssim 10^{-3}$, or with any asymmetric combination that leads to a similar rejection, we can achieve $<1$ background events. 
Hence, in the reminder of the discussion, our projections for the sensitivity assume a background-free experiment. Consequently, the 95\% CL region corresponds to $N_X = 3$\,.

The requirements on \Rsel can be translated to a general requirement for the detector performance.
For \Rsel the main properties that can be used to discriminate between signal and background are the azimuthal angles of the two photons, the decay vertex two photons, the transverse momentum of the diphoton system and the invariant mass. For neutron rejection in addition the calorimeter shower properties and the time of arrival can be explored. 
It can be shown that the required background reduction is achievable with existing detector (calorimeter) technologies: time resolution\footnote{The ultimate time resolution of the EM calorimeter will also determine the required cosmic muons veto strength and hence the required arrangement of muon chambers.} of $\sim\cO(10-100)~{\rm ps}$~\cite{Bornheim:20159Z, Lobanov_CMS_HGCAL:2020, Martinazzoli_LHCb:9006906}, energy resolution of a few percent and finally, position and angular resolutions of $\sim\cO(100)\,\mu{\rm m}$ and $\sim\cO(100)\,{\rm mrad}$ respectively~\cite{Bornheim:20159Z, Martinazzoli_LHCb:9006906, Liu_HGCalo:2020, Bonivento_SHiP_ecal:2018}.
To conclude this discussion, we note that even a subset of these specifications is sufficient to meet our minimal requirements for the search.

\begin{figure}[t]
	\includegraphics[width=0.45\textwidth]{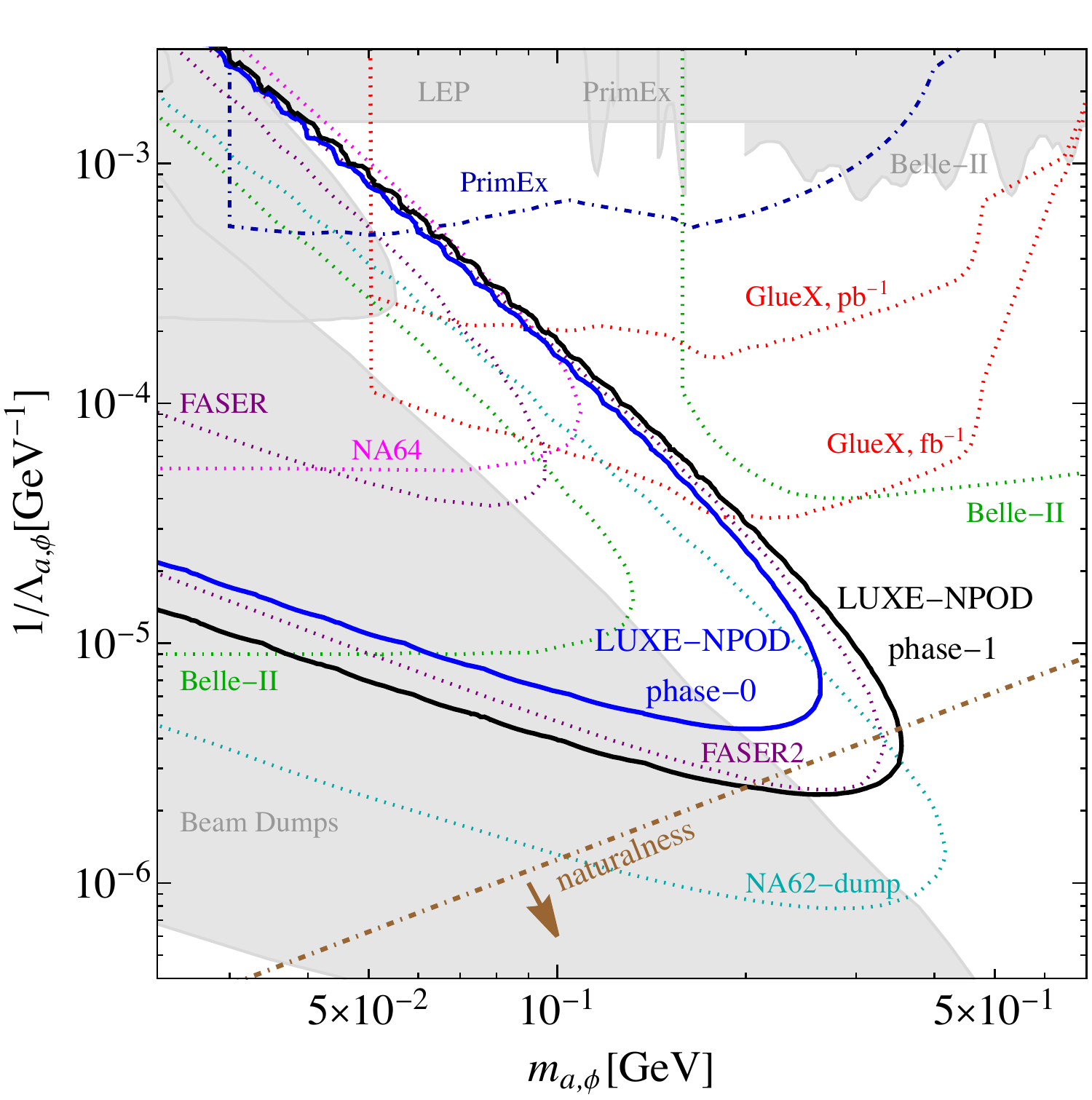}
	\caption{
	The projected reach of \LUXE-NPOD phase-0\,(1) in a solid blue\,(black) compare to the currently existing bounds (gray regions) on $X=a,\phi$-photon couplings from LEP~\cite{Jaeckel:2015jla,Knapen:2016moh,Abbiendi:2002je}, {\sc PrimEx}~\cite{Aloni:2019ruo}, Belle-II~\cite{BelleII:2020fag}, NA64~\cite{Dusaev:2020gxi,Banerjee:2020fue} and beam-dumps~\cite{Bjorken:1988as,Blumlein:1990ay,Blumlein:1991xh}.  
	The dark blue dot-dashed is the projection from on-tape {\sc PrimEx} dataset~\cite{Aloni:2019ruo}. 
	The dotted lines are future projections of NA62, Belle-II, FASER, {\sc PrimEx} and {\sc GlueX}~\cite{Dobrich:2015jyk,Dobrich:2019dxc,Dolan:2017osp,Feng:2018pew,Berlin:2018pwi,Aloni:2019ruo}.
	 The natural region for the scalar model is below the brown dashed-dotted line. }
	\label{fig:ReachSec}
\end{figure}

The projections for the sensitivity of  \LUXE-NPOD are shown in Fig.~\ref{fig:ReachSec} for the axion or scalar mass versus the effective coupling as defined in Eq.~\eqref{eq:Laphi}.
We compare the result to the current bounds from LEP~\cite{Jaeckel:2015jla,Knapen:2016moh,Abbiendi:2002je}, {\sc PrimEx}~\cite{Larin:2010kq,Aloni:2019ruo}, NA64~\cite{Dusaev:2020gxi,Banerjee:2020fue}, Belle-II~\cite{BelleII:2020fag}, and beam-dumps experiments~\cite{Bjorken:1988as,Blumlein:1990ay}.
In addition, the future projections of NA62 (in dump-mode), Belle-II, FASER, \PrimEx and \GlueX~\cite{Dobrich:2015jyk,Dobrich:2019dxc,Dolan:2017osp,Feng:2018pew,Aloni:2019ruo} are presented.
We see that already in phase-0 \LUXE can probe an unexplored parameter space in the mass range of $50\,\MeV\lesssim m_X \lesssim 250\,\MeV$ and $1/\Lambda_X > 4\times 10^{-6}\,\GeV^{-1}$. 
Moreover, we see that \LUXE phase-1 is expected to probe $40\,\MeV\lesssim m_X \lesssim 350\,\MeV$ and $1/\Lambda_X > 2\times 10^{-6}\,\GeV^{-1}$.
The region of natural parameter space for scalar is below the brown dashed-dotted line of Fig.~\ref{fig:ReachSec}, see Eq.~\eqref{eq:natu}, which will be probed in phase-1. 
The projections from the FASER2 (planned for a HL-LHC future run~\cite{Apollinari:2015wtw}) and NA62 in dump mode are roughly similar with the \LUXE\ phase-1 sensitivity curve, that is expected to be reached after one year of running.

An interesting comparison of the \LUXE-NPOD proposal presented here is with the case of electron beam-dump.
In this setup the electron beam is directly collided with the (same) dump.
For electron beam with $E_e=16.5$\,GeV the number of photons with $E_\gamma>1$\,GeV is $\sim 7$ per initial electron. 
Thus, naively, one can expect that the $X$ yield will be a factor of $\sim 2$ larger than in the \LUXE\ setup used as NPOD.
However, based on our \Geant simulation, the number of two-photon background events at the detector is expected to be much higher.
Particularly, we have estimated $\nbxonem{\gamma}$ and $\nbxonem{n}$ to be $\sim 20$ and $\sim 4$ and times larger than the respective values for the \LUXE-NPOD setup.
Hence, these estimations make the $X$ search much more challenging in terms of the requirements on the detector (\fngam\ and \Rsel).
The respective values of $R_{\gamma/n}$, \nbx{n}{1.0\mbox{~m}} and \nbx{\gamma}{1.0\mbox{~m}} for this setup are quoted in the Supplemental Material and the derivation of the probabilities is identical to the one of \LUXE-NPOD given above.

\section{Outlook}

In this work we propose a novel way to search for feebly interacting massive particles, exploiting striking properties of systems involving collision of high-energy electrons with intense laser pulses.
The laser medium acts effectively as a thick-material for electrons, which emit a large flux of hard collinear photons.
The same laser medium acts effectively as a thin-material
for these photons, which are free-streaming inside it.
The electron-laser collision is thus an apparatus, which efficiently convert UV electrons to a large flux of hard photons.

We then propose to direct this unique large and hard flux of photon onto a physical dump to allow the production of feebly interacting massive particles in a region of parameters never been probed before. 
We denote this apparatus as optical dump or NPOD (new physics search with optical dump).

This may seem like a pretty unique set of specifications to follow.
However, it happens to be that the proposed \LUXE experiment at the Eu.XFEL fulfils all the basic requirements of the above experimental concept.
\LUXE is a part of a broader worldwide program aiming to probe non-perturbative aspects of QED and field theories in general.
It is quite remarkable that even in its phase-0 and definitely in its phase-1, \LUXE will be able to probe uncharted territory of spin-0 feebly coupled particles.
This can be done in a nearly ``parasitic-mode'' of operation, where the only requirement is an additional detector system as proposed in this work, with no modification otherwise to the experimental design.
Moreover, we show that with a reasonable choice of detector technology, this search can be regarded as background-free. 

There are several future directions that would be also interesting to consider as followups to this project. 
The first is to investigate spin-0 particles that couple to gluons~\cite{Aloni:2019ruo} and types of non-linear photon dynamics~\cite{Bernard:1997kj,Evans:2018qwy,Bogorad:2019pbu,Gao:2020anb,Gorghetto:2021luj}.
Furthermore there are several experiments that aim to probe strong field QED.
It would be interesting if these would also consider adopting the NPOD concept to probe other part of the parameter space of our theories.
In that context, investigating other types of interactions or maybe even changing the flavor of the incoming particles would be also be extremely interesting.
Finally, it would  be interesting to explore what reach could be obtained when using parasitically the high-energy electron beams of future Higgs factories, e.g. the ILC~\cite{Baer:2013cma}, FCC-ee~\cite{FCC:2018evy}, CEPC~\cite{CEPCStudyGroup:2018ghi} and CLIC~\cite{Linssen:2012hp}.

\begin{acknowledgments}
The authors would like to thank Or Hen, Ben King, Andreas Ringwald, Mike Williams for useful discussions, Iftah Galon for help with \MG.
We are also grateful to Ben King, Federico Meloni and Mike Williams for comments on the manuscript. 
ZB thanks the Weizmann Institute of Science for hospitality via the Yutchun Program during the initial phase of this project.
OB and BH thank the DESY directorate for funding this work through the DESY Strategy Fund. 
The work by B.~Heinemann and was in part funded by the Deutsche Forschungsgemeinschaft under Germany‘s Excellence Strategy – EXC 2121 ``Quantum Universe" – 390833306. 
TM is supported by “Study in Israel” Fellowship for Outstanding Post-Doctoral Researchers from China and India by PBC of CHE.
The work of GP is supported by grants from 
BSF-NSF (No. 2019760), 
Friedrich Wilhelm Bessel research award, 
GIF, 
the ISF (grant No. 718/18),
Minerva, SABRA - Yeda-Sela - WRC Program, the Estate of Emile Mimran, and The Maurice and Vivienne Wohl Endowment.
The work of YS and TM is supported by grants from the NSF-BSF (No. 2018683), by the ISF (grant No. 482/20) and by the Azrieli foundation. 
YS is Taub fellow (supported by the Taub Family Foundation).
The work of NTH is supported by a research grant from the Estate of Dr. Moshe Gl\"{u}ck, the Minerva foundation with funding from the Federal German Ministry for Education and Research, the ISF (grant No. 708/20), the Anna and Maurice Boukstein Career Development Chair and the the Estate of Emile Mimran.
Simulations of photon production were enabled by resources provided by the Swedish National Infrastructure for Computing (SNIC) at the High Performance Computing Centre North (HPC2N), partially funded by the Swedish Research Council through grant agreement no. 2018-05973.
\end{acknowledgments}

\twocolumngrid
\bibliographystyle{utphys}
\bibliography{LUXE_BSM}

\clearpage
\newpage
\maketitle
\onecolumngrid

\begin{center}
\textbf{\large Probing new physics at the non perturbative QED frontier} \\
\vspace{0.05in}
{ \it \large Supplemental Material}\\
\vspace{0.05in}
{Zhaoyu Bai, Thomas Blackburn, Oleksandr Borysov, Oz Davidi, Anthony Hartin, Beate Heinemann, Teng Ma, Gilad Perez,  Arka Santra, Yotam Soreq, and Noam Tal Hod}
\end{center}

\onecolumngrid
\setcounter{equation}{0}
\setcounter{figure}{0}
\setcounter{table}{0}
\setcounter{section}{0}
\setcounter{page}{1}
\makeatletter
\renewcommand{\theequation}{S\arabic{equation}}
\renewcommand{\thefigure}{S\arabic{figure}}
\renewcommand{\thetable}{S\arabic{table}}
\newcommand\ptwiddle[1]{\mathord{\mathop{#1}\limits^{\scriptscriptstyle(\sim)}}}

\section{Secondary New Physics production of spin-0 particles}

\begin{figure*}[t]
    \includegraphics[width=0.4\textwidth]{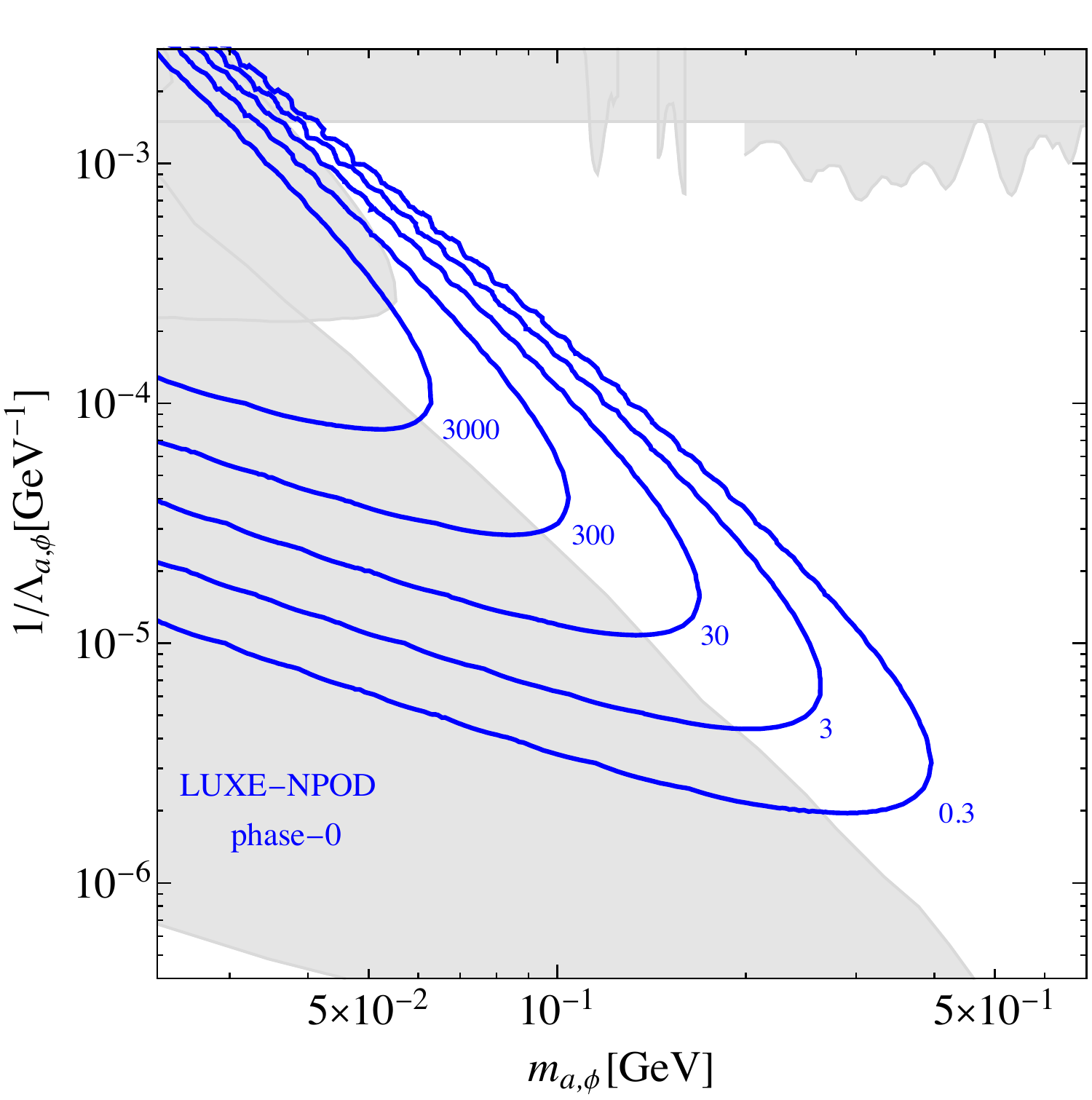}~~~
	\includegraphics[width=0.4\textwidth]{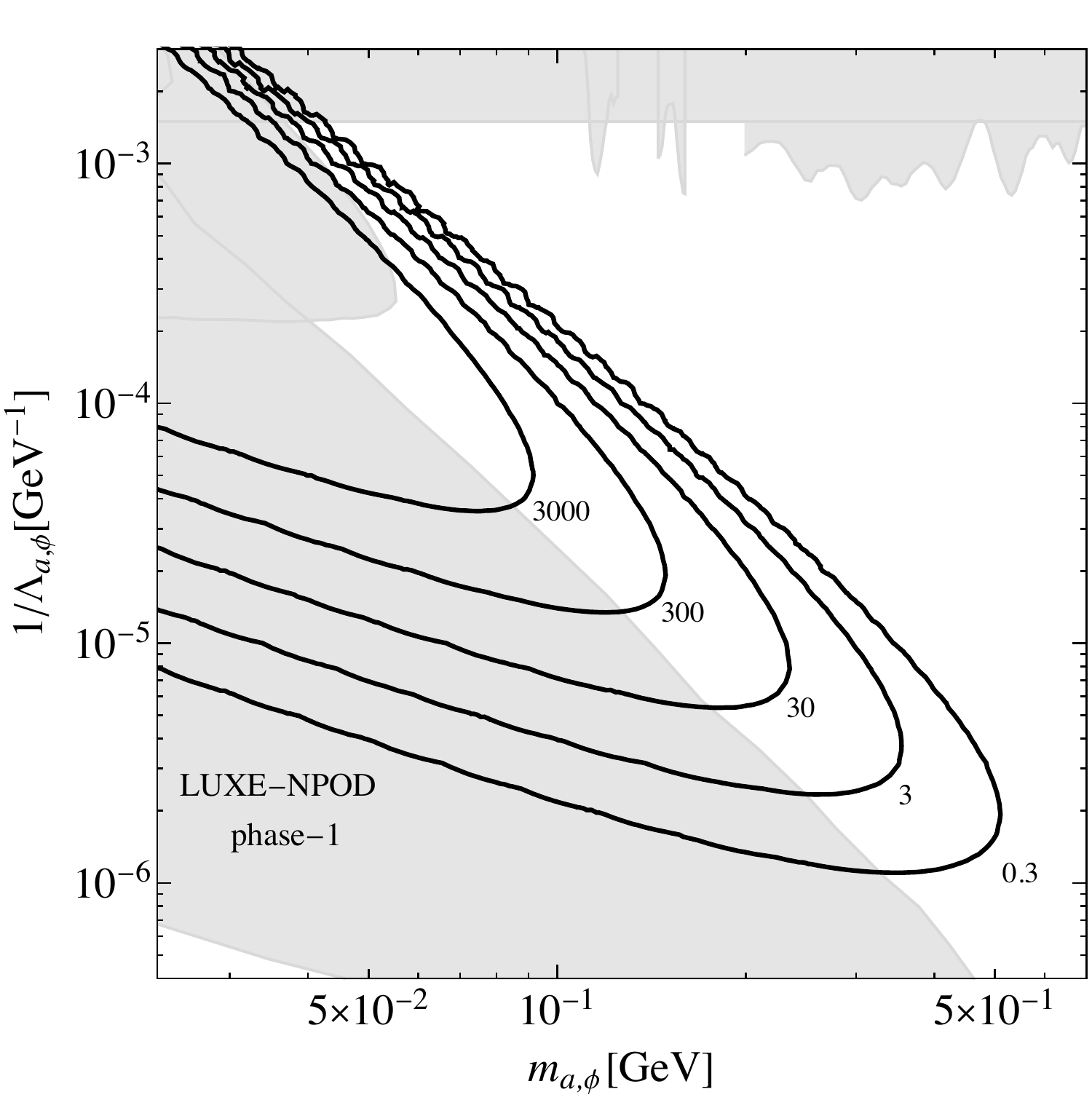}
		\caption{
	Contours of the expected number of $X$ events, $N_X$, for phase-0\,(phase-1) on the left\,(right) panel.  
	}
	\label{fig:NaConoutrs}
\end{figure*}

Contours of the expected number of spin-0 signal events in secondary production are showed in Fig.~\ref{fig:NaConoutrs}. 
The signal estimation is described in the main text by using Eq.~\eqref{eq:NaSec} and was verified by a \MG simulation. 

\section{Primary New Physics production}

\begin{figure*}[t]
    \includegraphics[width=0.28\textwidth]{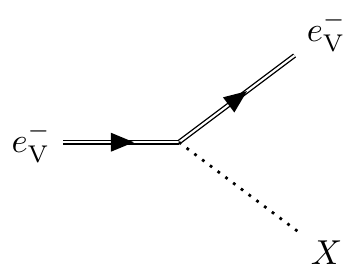}~
	\includegraphics[width=0.28\textwidth]{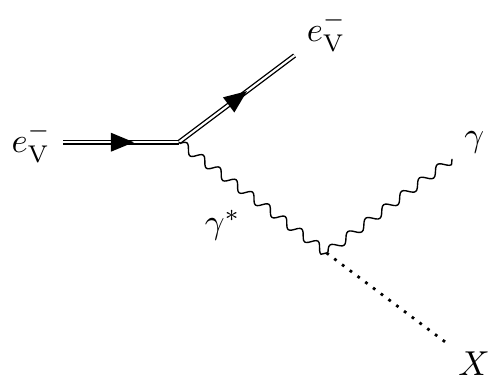}~
	\includegraphics[width=0.28\textwidth]{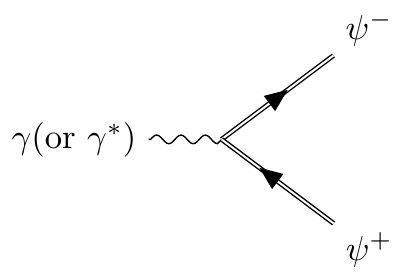}
	\caption{
	The Feynman diagrams for primary new physics production. 
	The dressed electron is represented as a double line. 
	Left: $\eVol \to \eVol + X$, see Eq.~\eqref{eq:aComptonProd};
	middle: $\eVol \to \eVol + \gamma^* \to \eVol + X+ \gamma $, see Eq.~\eqref{eq:aOffshellProd};
	right: $\gamma \, (\text{or }\gamma^*) \to \psi^+ + \psi^-$, see Eq.~\eqref{eq:mCPprod}.
	}
	\label{fig:PrimDiag}
\end{figure*}

In this section we calculate new physics primary production of the processes in Eqs.~\eqref{eq:aComptonProd}--\eqref{eq:mCPprod}, namely 
\begin{align}
    \eVol \to \eVol + X \, , \quad
    \eVol \to \eVol + \gamma^* \to \eVol + X + \gamma \, , \quad
    \gamma \, (\text{or }\gamma^*)  \to \psi^+ + \psi^- \, .
\end{align}
The Feynman diagrams of the above processes are plotted in Fig.~\ref{fig:PrimDiag}.  
The ratio between the above production rates and the QED-only Compton scattering for $\xi=3.4$ and $\chi=0.65$ are plotted in Fig.~\ref{fig:PrimDiag}, where we set $g_{ae}=10^{-8}$, $\Lambda_a=\TeV$ and $q=5\times 10^{-5}$. 
Our calculation below are based on Ref.~\cite{Ritus1985}.

\begin{figure}[t]
	\includegraphics[width=0.5\textwidth]{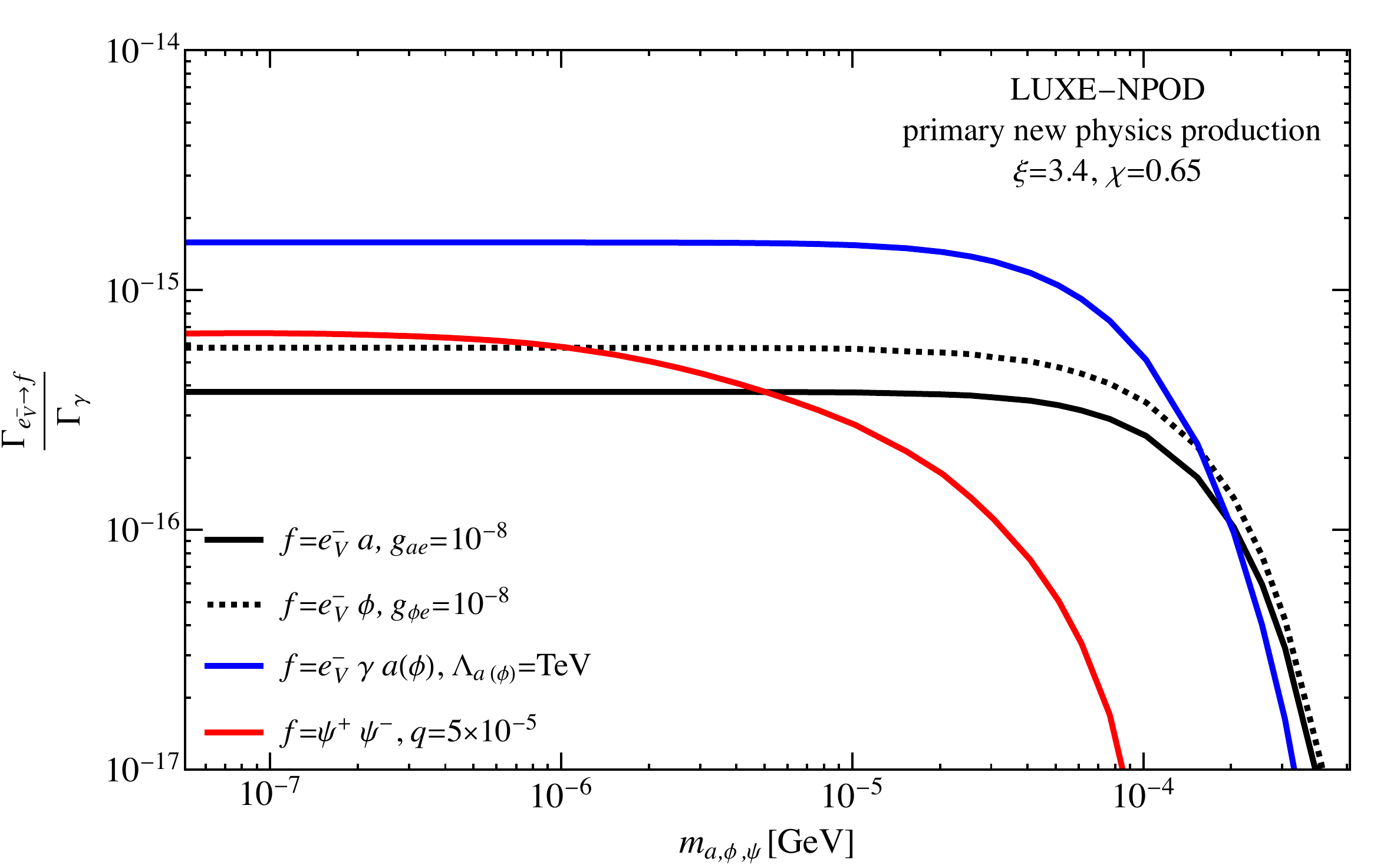}
	\caption{The new physics primary production rates of processes in Eqs.~\eqref{eq:aComptonProd}--\eqref{eq:mCPprod} for $\xi=3.4$ and $\chi=0.65$ normalized to the photon production rate, $\taugam=1/\Gagam\approx12\,\fs$. }
	\label{fig:PrimaRate}
\end{figure}

\subsection{Non perturbative $X$ production in a circularly polarised laser}

We calculate the non-perturbative Compton emission of ALP and scalar in a circular polarized laser. 
The background laser field can be written as
\begin{align}
    A_\mu 
=   a_{1\mu} \cos(k\cdot x) +a_{2\mu} \sin(k \cdot x) \, ,
\end{align}
where $k_\mu$ is the laser four vector, $x_\mu$ is the spatial coordinate, $A^2 \equiv a_1^2 = a_2^2$ and $k\cdot a_{1,2}=0$.
The solution of the Dirac equation for electron with momentum $p_\mu$ ($p^2=m^2_e$) in the above laser background is given by the Volkov state~\cite{Volkov} 
\begin{align}
    \psi_{pr}(x)
=&  \left( 1+\frac{e\slashed{k}\slashed{A}}{2kp}\right) \,u_{r}(p)\,
    \exp\,\left[ -ipx -i\int^{kx}_0d\phi e\left(\frac{pA}{kp}-\frac{e A^2}{2kp}\right) \right]\\
=&  \left[1+\frac{e\slashed{k}\slashed{a}_{1}}{2kp}\sin(kx)
    +\frac{e\slashed{k}\slashed{a}_{2}}{2kp}\cos(kx) \right]
    \,u_{r}(p)\,\exp\,\left[-iqx -ie\left( \frac{a_{1}p}{kp}\sin(kx)+\frac{a_{2}p}{kp}\cos(kx)\right)\right]\, ,
\end{align}
where $u_r(p)$ is the Dirac spinor in free space. 
For convenience, we define the followings 
\begin{align}
    q_\mu \equiv p_\mu+\frac{e^2a^2}{2(kp)}k_\mu \, , \qquad 
    \xi = \frac{e a}{m_e} \, , \qquad
    \chi \equiv \frac{kq}{m_e^2}\xi = \frac{kp}{m_e^2}\xi    \, , \qquad
    u \equiv \frac{k\cdot p_a}{k\cdot q'} \, , \qquad
    r_m \equiv \frac{m_a}{m_e} \, ,
\end{align}
such that $m_*^2 \equiv q^2 = m_e^2(1+\xi^2)\,$.

The $\eVol(p) \to \eVol(p') a(p_a) $ amplitude can be written as 
\begin{align}
    \cM_{e\to e a}
=   g_{ae}\int\frac{\text{d}^4x}{\sqrt{2^3p_{a,0}q_0q'_0}}\bar{u}_{r'}(p')
    &\left[
     \gamma_5+ \frac{4\beta}{ea^2}\slashed{k}\slashed{a}_1\gamma_5\sin (kx)
    +\frac{4\beta}{ea^2}\slashed{k}\slashed{a}_2\gamma_5\cos(kx)
    \right]
    \nonumber\\
     &\times u_{r}(p) e^{-i(\alpha_1 \text{sin}kx-\alpha_2\text{cos}kx)+i(q-q'-p_a)x} \, ,
\end{align}
where 
\begin{align}
    \alpha_{1,2} 
=   -e \frac{a_{1,2}\cdot p'}{k\cdot p'} \, \quad
    \beta 
=   \frac{e^2a^2}{8}\left(\frac{1}{k\cdot q} - \frac{1}{k \cdot q'} \right)
=   -\frac{u \xi^3}{8\chi_e}
\end{align}
The exponent can be expanded into a Bessel functions as
\begin{align}
    (1,\cos\phi,\sin\phi)e^{-iz\sin(\phi-\phi_0)}
=   \sum^\infty_{s=-\infty}(B_1,B_2,B_3)e^{-is\phi}.
\end{align}
with
\begin{align}
    B_1
=   J_s(z)e^{-is\phi_0} \, , \quad
    B_2
=   \frac{1}{2}\left[J_{s-1}(z)e^{-i\phi_0}+J_{s+1}e^{i\phi_0}\right]
    e^{is\phi_0}\, , \quad
    B_3
=   \frac{1}{2i}\Big[J_{s-1}(z)&e^{-i\phi_0}-J_{s+1}e^{i\phi_0}\Big]e^{is\phi_0} \, ,
\end{align}
and $z=\sqrt{\alpha_1^2+\alpha_2^2}$ and $\tan\phi_0=-\alpha_1/\alpha_2\,$.
The ALP emission rate is evaluated by squaring the amplitude and evaluating the phase space integration, which is straightforward in the center of mass system. 
The resulting rate is
\begin{align}
    \Gamma_{e\to e a}
=   -\frac{g_{ae}^2m_e^2}{8\pi q^0} \sum_{s>s_0}\int^{u_2}_{u_1}\frac{\text{d}u}{(1+u)^2}
    \Bigg\{&
    \left[\frac{r_m^2}{2}
    +\xi^2\left(\frac{1+(1+u^2)}{2(1+u)}-\frac{s\chi}{\xi^3(1+u)}-1\right)\right]J_s^2(z)
    \nonumber \\
    &-\frac{\xi^2}{4}\frac{u}{1+u}\left[J_{s-1}^2(z)+J_{s+1}^2(z)\right] 
    \Bigg\} \, , 
\end{align}
where 
\begin{align}
    z
=&  \frac{\xi^2}{\chi}\sqrt{\bigg(\frac{2s\chi}{\xi}-r_m^2\bigg)u-(1+\xi^2)u^2-r_m^2} \, , \\
    s_0 
=&   \frac{r_m \xi}{\chi_e} \left( r_m + 2\sqrt{1+\xi^2} \right) \, ,  \\
    u_{1,2}
=&  \frac{s\chi}{\xi(1+\xi^2)} - \frac{r_m^2}{2(1+\xi^2)}
    \pm\sqrt{\bigg(\frac{s\chi}{\xi(1+\xi^2)}-\frac{r_m^2}{2(1+\xi^2)}\bigg)^2-\frac{r_m^2}{1+\xi^2}}\, .
\end{align}

The production rate of scalar with electron coupling is $g_{\phi e}\phi\bar{e}e$ via the process of $\eVol\to \eVol + \phi$ is calculated similary to the case of ALP.
The rate is given by
\begin{align}
    \Gamma_{e\to e \phi}
=   \frac{g_{\phi e}^2m_e^2}{8\pi q^0} \sum_{s>s_0}\int^{u_2}_{u_1}\frac{\text{d}u}{(1+u)^2}
    \Bigg\{&
    \left[2-\frac{r_m^2}{2}
    -\xi^2\left(\frac{1+(1+u^2)}{2(1+u)}-\frac{s\chi}{\xi^3(1+u)}-1\right)\right]J_s^2(z)
    \nonumber \\
    &+\frac{\xi^2}{4}\frac{u}{1+u}\left[J_{s-1}^2(z)+J_{s+1}^2(z)\right] 
    \Bigg\} \, , 
\end{align}
%

\subsection{Off-shell ALP production in a circularly polarised laser}

The amplitude for the process $\eVol(p)\to \eVol(p^\prime) + \gamma (k^\prime) +a (p_a)$ or $\eVol(p)\to \eVol(p^\prime) + \gamma (k^\prime) +\phi (p_\phi)$, can be written as
\begin{align}
    \cM_{e\to e \gamma a\,(\phi)} 
=   \frac{e}{l^2} \sum_{s} \bar{u}_{r^\prime}(p^\prime)
    \left\{
    \left[ \slashed G \!+\!\frac{e^2 A^2 (k\cdot G) \slashed k }{2(k\cdot p)(k\cdot p^\prime) } \right] C_0 
    \!+e\left[ \frac{\slashed a_1 \slashed k \slashed G }{2k\cdot p^\prime}  
    \!+\! \frac{\slashed G \slashed k  \slashed a_1   }{2k\cdot p} \right] C_1 
    \!+e\left[ \frac{\slashed a_2 \slashed k \slashed G }{2k\cdot p^\prime}  
    \!+\! \frac{\slashed G \slashed k  \slashed a_2}{2k\cdot p} \right] C_2  
    \right\} u_r(p)\,,  
\end{align}      
where $l=k^\prime +p_a$, 
$G_\nu= \frac{l^\mu k^\prime_\rho \epsilon^\ast_\sigma(k^\prime) \epsilon^{\mu \nu \rho \sigma} }{\Lambda_a}\, \left(G_\nu = \frac{l\cdot k^\prime \epsilon^\ast_\nu(k^\prime)-l\cdot \epsilon^\ast_\sigma(k^\prime)k^\prime_\nu }{\Lambda_\phi}\right)$ parametrises the ALP-photon\,(scalar-photon) interaction and
\begin{align}
    C_0(s \alpha_1\alpha_2) 
=&  J_s(z)e^{-is\varphi} \, , \\
    C_1(s \alpha_1\alpha_2)
=&  \left[\frac{s}{z} J_s(z) \cos \varphi +i J_s^\prime(z) \sin \varphi \right]e^{-is\varphi} \, ,  \\
    C_2(s \alpha_1\alpha_2)
=&  \left[ \frac{s}{z} J_s(z) \sin \varphi -i J_s^\prime(z) \cos \varphi \right]e^{-is\varphi}\, ,
\end{align}
with $\varphi$ is the azimuthal angle of the outgoing electron, 
\begin{align}
    \alpha_i 
=   e\left(\frac{a_i \cdot p}{k \cdot p} -\frac{a_i \cdot p^\prime}{k \cdot p^\prime}  \right)\, , \quad 
    z
=   \sqrt{\alpha_1^2+\alpha_2^2}
=   \frac{\xi^2}{\chi}\sqrt{(1+u)(\lambda_2-\lambda)}\, , \quad 
    \cos \varphi 
=   \frac{\alpha_1}{z}\, \quad
    J^\prime_s(z) 
=   \frac{d J_s(z)}{d z} \,.
\end{align}  

The ALP production rate per unit time and unit volume can be obtained 
\begin{align}
    W 
=  \frac{1}{2VT} \sum_{r r^\prime} 
    \int \frac{d^3p_a d^3 k^\prime d^3 q^\prime}{(2\pi)^9 (2E_a)(2E_\gamma) (2 q^\prime_0)   }
    \left|\cM_{e\to e \gamma a\,(\phi)} \right|^2 (2\pi)^4\delta^{(4)}(sk+q-q^\prime -l) \, .
\end{align}
We integrate over the ALP and photon phase space and express the outgoing electron phase space integration $d^3 q^\prime$ in terms of new variables $u =k\cdot l/k\cdot q^\prime$ and $\lambda=l^2/m_e^2$.
Therefore, we get that the ALP off-shell production rate per initial electron per unit volume is.
\begin{align}
    \Gamma_{e\to e \gamma a\,(\phi)}
=   \frac{2\pi \alpha  m_e^2}{q_0 \Lambda^2_{a\,(\phi)}} \sum_{s> s_0} \int \frac{d^3 q^\prime}{q_0^\prime} w 
=   \frac{2\pi^2 \alpha  m_e^4}{q_0 \Lambda^2} \sum_{s> s_0}  
    \int_{u_1}^{u_2} \frac{d u}{(1+u)^2}
    \int_{\lambda_1}^{\lambda_2}  d \lambda w \, ,
\end{align}   
where
\begin{align}
    w
=&  \frac{ 4 \left(s^2J_s^2+ z^2 J_s^{\prime 2} \right) \lambda \xi^4 \left(u^2+2u +2\right) -4z^2 J_s^2  X}{\lambda^2 \xi^2 (u+1)z^2} 
    \frac{(\lambda-\lambda_1)^3}{3072 \pi^4 \lambda^2 }\, ,  \\ 
    X 
=&   \lambda^2 \xi^2(u+1)  -8 s^2 \chi_e^2 +\lambda \xi \left(\xi^3(u^2+2u+2) +2\xi(u+1) -4s u\chi_e\right)\, , \\ 
    s_0 
=&  \frac{(m_a +m_e^{\ast})^2 -m_e^{\ast 2}}{2 k\cdot q}
=   \frac{r_m \xi}{2\chi}\left(r_m +2\sqrt{1+\xi^2}\right)\, ,  
 \end{align}
where the Bessel functions argument is $z$, $E_s^2 =(q+s k)^2=m_e^2(\xi^3+\xi+2s\chi_e)/\xi$ and
\begin{align}
    u_{2,1} 
=&  \frac{E_s^2 -m_a^2 -m_e^{\ast 2}\pm\sqrt{(E_s^2 -m_a^2 -m_e^{\ast 2})^2 -4m_a^2 m_e^{\ast 2} } }
    {2 m_e^{\ast 2} } \nonumber\\
=&   \frac{2s\xi-r_m^2 \xi \pm \sqrt{r_m^4\xi^2+4s^2\chi_e^2-4r_m^2\xi(\xi+\xi^3+s\chi_e)}}{2\xi(1+\xi^2)}\, ,  \\
    \lambda_1 
=&  r_m^2\, , \quad
    \lambda_2 
=   \frac{E_s^2 u}{m_e^2(1+u)} -(1+\xi^2) u \, .
\end{align}
%

\subsection{Millicharged particle production in a circularly polarized laser}

Equation~\ref{eq:mCPprod} describes the direct production of millicharged particle (mCP) pairs by high-energy photons in a strong laser background: see also the right-hand panel of Fig.~\ref{fig:PrimDiag}.
Here we consider the case where the high-energy photons are produced by nonlinear Compton scattering, i.e. where the photons are emitted and subsequently decay within the same laser pulse.
The rate (per unit time) at which mCP pairs are produced by nonlinear Compton photons in an ultraintense electromagnetic (EM) wave with invariant amplitude $\xi$ and wavevector $\kappa$, is given by:
\begin{align}
    \Gamma_\mathrm{mCP} 
=   \frac{1}{\Gagam} \int_0^1 \Gamma_\pm(s \eta_e, \xi, q, r_m) 
    \frac{d\Gagam(\eta_e, \xi)}{d s} \, d s,
\end{align}
where $\Gamma_\pm(\eta_\gamma, \xi, q, r_m)$ is the rate at which a photon with energy parameter $\eta_\gamma = \kappa \cdot k / m^2_e$ (momentum $k$) creates a pair of particles with charge and mass ratio $q$ and $r_m=m_\psi/m_e$, $\Gagam(\eta_e, \xi)$ is the rate at which an electron with energy parameter $\eta_e = \kappa \cdot p / m^2_e$ (momentum $p$) emits photons, and $s = \eta_\gamma / \eta_e$.
(The quantum parameter is recovered as $\chi_e = \xi \eta_e$.)

We use the rates as calculated in the locally monochromatic approximation~\cite{Heinzl:2020ynb,Blackburn:2021rqm}, which assumes the background EM field is a plane wave.
For the photon emission rate, we have
\begin{align}
    \frac{d \Gagam(\eta_e, \xi)}{d s} 
=   -\frac{\alpha m^2}{p^0} \sum_{n=1}^\infty
    \left\{ J_n^2(z) + \frac{\xi^2}{2} \left[ 1 + \frac{s^2}{2(1-s)} \right]
    \left[ 2 J_n^2(z) - J_{n-1}^2(z) - J_{n+1}^2(z) \right]
    \right\},
\end{align}
where the bounds on $s$, for each $n$, are $0 < s < s_n / (1 + s_n)$, and the auxiliary variables are
\begin{align}
    z^2 
=  \frac{4 n^2 \xi^2}{1 + \xi^2}\frac{s}{s_n (1-s)}
    \left[ 1 - \frac{s}{s_n (1-s)} \right]\, , \quad
    s_n 
=  \frac{2 n \eta_e}{1 + \xi^2} \, .
\end{align}
For the mCP pair creation rate, we have
\begin{align}
    \label{eq:PairCreationRate}
    \frac{d \Gamma_\pm (\eta_\gamma, \xi, q, r_m)}{d s} 
=   q^2 r_m^2 \frac{\alpha m^2}{k^0} \sum_{n=n_\star}^\infty
    \left\{ J_n^2(z) - \frac{\xi^2}{2}
    \left[ \frac{1}{2 s (1 - s)} - 1 \right]
    \left[ 2 J_n^2(z) - J_{n-1}^2(z) - J_{n+1}^2(z) \right]
    \right\} \, ,
\end{align}
where $n_\star = \lceil 2 r_m^2 (1 + q^2 \xi^2/ r_m^2) /\eta_\gamma \rceil$, $\frac{1}{2} \left[1 - (1 - 4/s_n)^{1/2}\right] < s < \frac{1}{2} \left[1 - (1 - 4/s_n)^{1/2}\right]$, and the auxiliary variables are
\begin{align}
    \label{eq:PairCreationRateHelpers}
    z^2 
=  \frac{4 n^2 q^2 \xi^2 / r_m^2}{1 + q^2 \xi^2 / r_m^2}
    \frac{1}{s_n s (1-s)}
    \left[ 1 - \frac{1}{s_n s (1-s)} \right] \, ,
    s_n 
=   \frac{2 n \eta_\gamma / r_m^2}{1 + q^2 \xi^2 / r_m^2} \, .
\end{align}
Equations~\ref{eq:PairCreationRate} and \ref{eq:PairCreationRateHelpers} are obtained from the electron-positron pair creation rates by making the following transformations: $\alpha \to q^2 \alpha$, $m_e \to r_m m_e$, $\xi \to q \xi / r_m$, $\eta \to \eta / r_m^2$.
The result of the calculation is $\Gamma_\mathrm{mCP}$ as a function of $\xi$ and $\eta_e$, for given charge and mass fraction $q$ and $r_m$.

\section{Background estimation}

The background estimation discussed in section~\ref{sec:npodproposal} relies on the assumption that for a fixed energy threshold, the ratio of the number of photons to the number of neutrons which arrive at the detector face per bunch crossing, $R_{\gamma/n}=\lambda_\gamma(L_D)/\lambda_n(L_D)$, is approximately constant for different $L_D$ values.

This assumption is important due to the extreme computational difficulty to fully simulate enough bunch crossings to allow a reliable estimation of the number of two-photon events with $E>0.5$~GeV at $L_D=1$~m.
In that case, more than $\sim 10^7$ bunch crossings would be needed.
We leave this extreme simulation campaign to a future work and instead, we rely on the constant $R_{\gamma/n}$ assumption as discussed below.

\begin{figure}[!th]
	\includegraphics[width=0.45\textwidth]{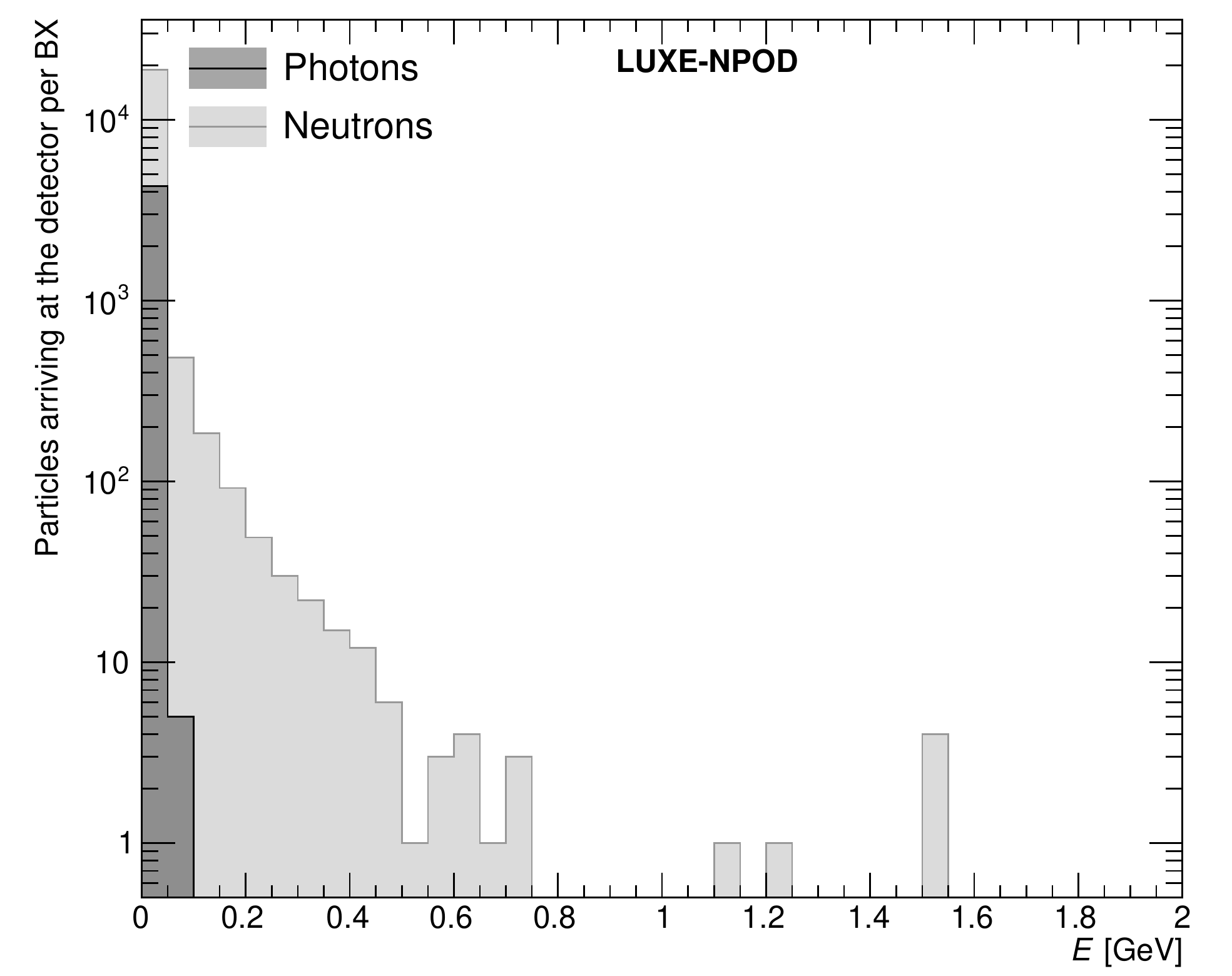}
	\includegraphics[width=0.45\textwidth]{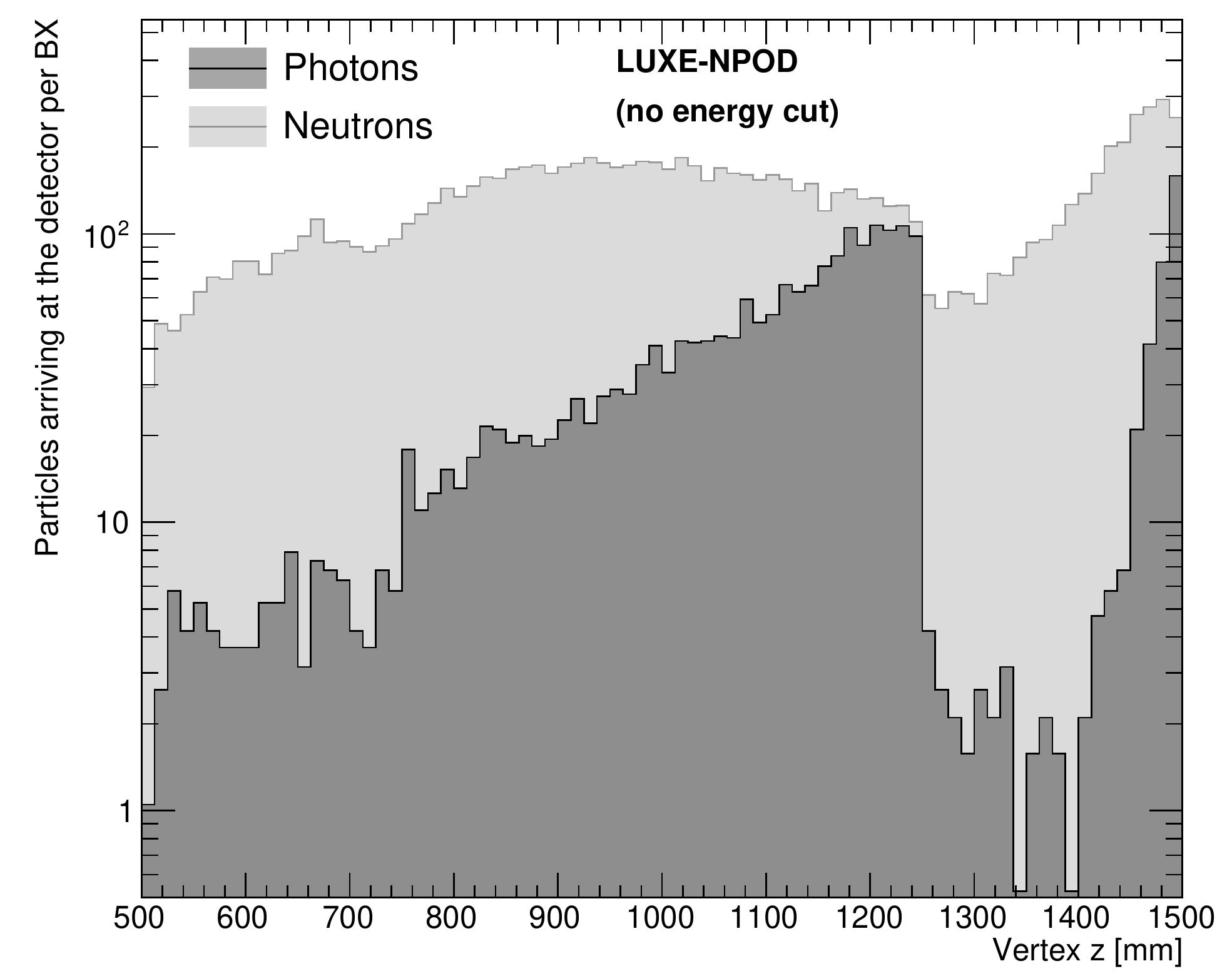}
	\caption{
	The energy (left) and the production vertex $z$-coordinate (right) of all background neutrons and photons arriving at the detector face for the \LUXE-NPOD setup. The two distributions are given with no energy cut. The production vertex distribution is only shown for the range of the 1~m long dump. The data in the two distributions correspond to two fully simulated bunch crossings.
	}
	\label{fig:bkg_E_vz}
\end{figure}

This assumption is in turn based on the observed correlation between the production of photons to that of neutrons, as can be seen in Fig.~\ref{fig:bkg_E_vz}.
The data shown in the figure is taken from a full \Geant v~10.06.p01~\cite{Agostinelli:2002hh,Allison:2006ve,Allison:2016lfl} simulation (using the ${\rm QGSP\_BERT}$ physics list) of two bunch crossings for the nominal \LUXE-NPOD setup discussed in Section~\ref{sec:npodproposal}.
Altogether there are $10^{10}$ primary photons distributed in energy as in Fig.~\ref{fig:dNdE} (for phase-1).
These primary photons are shot on the dump and produce electromagnetic and  hadronic showers of particles, which may escape the dump volume.
The kinematic properties of all particles which escape the dump and arrive at the detector face are saved for further analysis.
Since the $E>0.5$~GeV requirement leaves zero photons and only a handful of neutrons for $L_D=1$~m, this requirement is relaxed in order to plot the distributions seen in Fig.~\ref{fig:bkg_E_vz}.
It can be seen (left plot) that the energy of most of the particles is well below the photon's minimum energy of 0.5~GeV used in this work.
While the correlation between the trends of the two productions cannot be trivially explained within the scope of this work, it is clearly evident in the figure.
It can be expected that the correlation is independent of the particles minimum energy, as long as one looks at samples obtained using the same minimum energy for the two species (photons and neutrons).

\begin{figure}[!th]
	\includegraphics[width=0.45\textwidth]{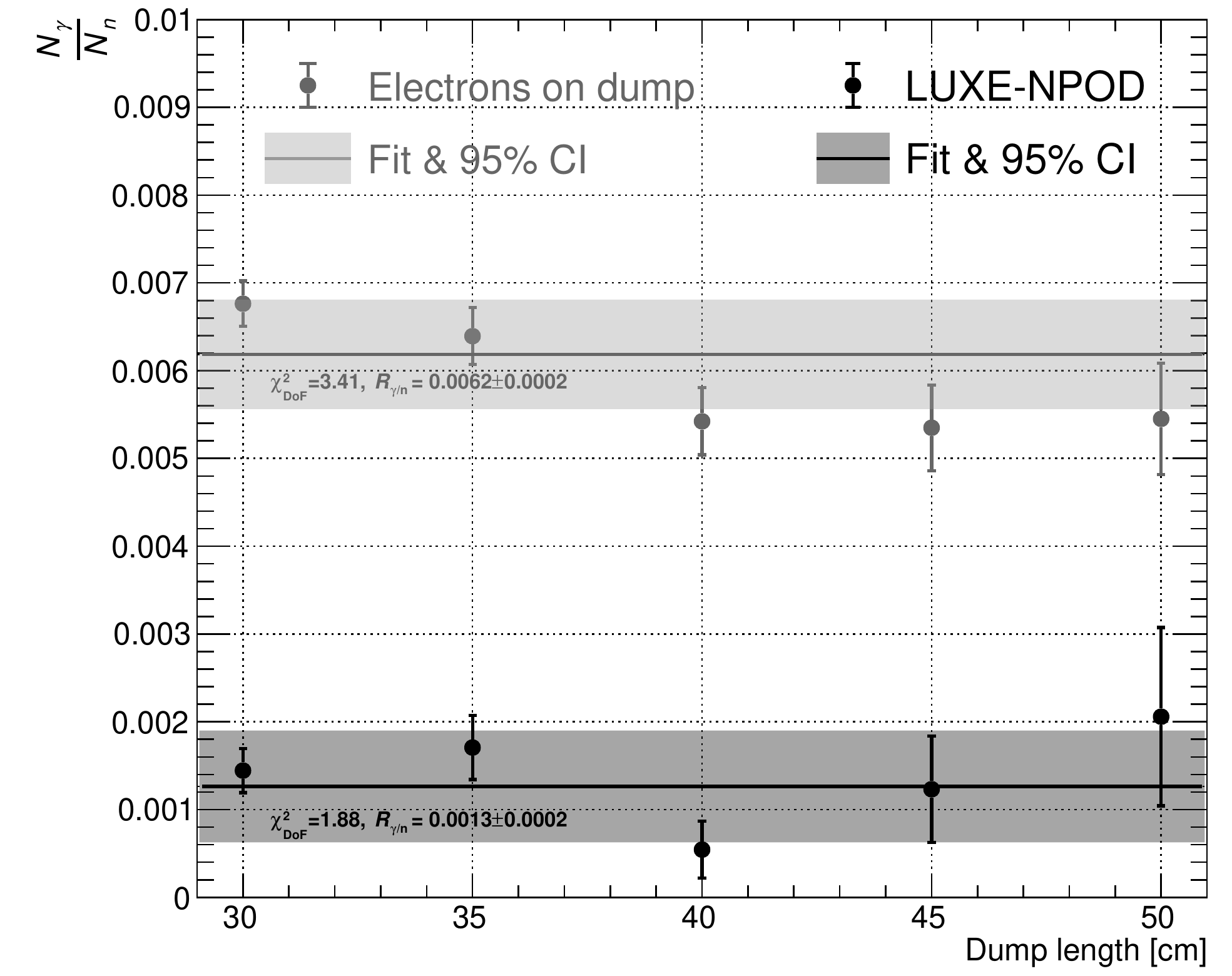}
	\caption{
	The ratio of number of photons to the number of neutrons vs different dump lengths ranging from 30~cm to 50~cm for the \LUXE-NPOD setup in black. For comparison, the same ratio is given for a hypothetical ``electrons on dump'' setup in grey. Each data point corresponds to a full simulation of two bunch crossings for the given dump length. Beyond 50~cm, there are only a few or no photons left and hence these points are not simulated. The fits to a zeroth order polynomial of the ratios are shown along with the corresponding 95\% confidence intervals. The fitted ratio, $R_{\gamma/n}$, is used to derive the probabilities to find two photons (real or fake) as discussed in the main text. 
	}
	\label{fig:bkg_Rgn}
\end{figure}

We exploit this observed correlation by studying the ratio of the two production rates for different dump lengths.
Particularly, we replicate the simulation of the same \LUXE-NPOD setup with the same statistics as discussed above for five dump lengths from 30~cm to 50~cm in steps of 5~cm, where in all cases the distance between the rear of the dump and the face of the detector is kept fixed at 2.5~m.
The statistics for dump lengths greater than 50~cm are too low and hence these points are not simulated.
For each point, the ratio of the number of photons to neutrons per bunch crossing is calculated for $E_{\gamma,n}>0.5$~GeV.
The points are shown in Fig.~\ref{fig:bkg_Rgn} together with a fit to a zeroth order polynomial from which we determine $R_{\gamma/n}$ for further analysis as discussed in section~\ref{sec:npodproposal}. 
This figure also includes the respective case of the ``electrons on dump'' hypothetical setup for the same dump lengths and otherwise the same conditions as discussed above.
In these cases, the only difference is that instead of simulating $10^{10}$ primary photons distributed in energy as in Fig.~\ref{fig:dNdE}, we simulate $3\times 10^9$ monochromatic ($E_e = 16.5$~GeV) primary electrons.

It can be seen that the ratio $R_{\gamma/n}$ can indeed be described as flat vs $L_D$, which indeed allows the extrapolation to $L_D=1$~m as discussed in section~\ref{sec:npodproposal}.
Furthermore, the resulting values for the two setups are significantly different with the ``electrons on dump'' result being much larger than the \LUXE-NPOD one ($0.0062\pm 0.0002$ vs $0.0013\pm 0.0002$ respectively).
This difference implies that while the two-photon background induced by real photos or neutrons faking photons will be manageable for the \LUXE-NPOD setup, it will be non-manageable for the hypothetical ``electrons on dump'' setup.

For completeness, we also quote the values of $\nbxonem{n}$ and $\nbxonem{\gamma}$ resulting from a simulation of two BXs, similarly to what is done for the \LUXE-NPOD setup.
These are found to be $\nbx{n}{1.0}=42.6\pm 4.6$ and $\nbx{\gamma}{1.0}=0$ for the ``electrons on dump'' case.
The derivation of the different background components is identical to the one given for the \LUXE-NPOD setup.

\end{document}